\documentstyle[aps,amssymb]{revtex}

\begin{document}
\title{Correct path-integral formulation of the quantum thermal field theory in the
coherent state representation }
\author{Su Jun-Chen and Zheng Fu-Hou}
\address{Center for Theoretical Physics, Physics College, Jilin University,\\
Changchun 130023, People's Republic of China}
\date{}
\maketitle

\begin{abstract}
The path-integral quantization of thermal scalar, vector and spinor fields
is performed newly in the coherent-state representation. In doing this, we
choose the thermal electrodynamics and $\varphi ^4$ theory as examples. By
this quantization, correct expressions of the partition functions and the
generating functionals for the quantum thermal electrodynamics and $\varphi
^4$ theory are obtained in the coherent-state representation. These
expressions allow us to perform analytical calculations of the partition
functions and generating functionals and therefore are useful in practical
applications. Especially, the perturbative expansions of the generating
functionals are derived specifically by virtue of the stationary-phase
method. The generating functionals formulated in the position space are
re-derived from the ones given in the coherent-state representation.


\end{abstract}

\section{Introduction}

In the quantum statistics, it has been shown that the partition functions
and thermal Green's functions for many-body systems may conveniently be
calculated in the coherent-state representation [1-6]. This is because the
partition function either for a boson system or for a fermion system can be
given a path (or say, functional) integral expression in the coherent-state
representation and, furthermore, one can write out a generating functional
of Green's functions for the system in the coherent-state representation
[1-6]. However, the path integral expression derived in the previous
literature can only be viewed as a formal symbolism because in practical
calculations, one has to return to its original discretized form which leads
to the path integral expression. If one tries to perform an analytical
calculation of the path integrals by employing the general methods and
formulas of computing functional integrals, one would get a wrong result.
This implies that the previous path integral expressions for the partition
function and the generating functional of Green's functions were not given
correctly. The incorrectness is due to that in the previous path integral
expressions, the integral representing the trace is not separated out and
the time-dependence of the integrand in the remaining part of the path
integral is given incorrectly. The partition function and the generating
functional of Green's functions were rederived in the coherent-state
representation and given correct functional-integral expressions in the
author's previous paper [7]. These expressions are consistent with the
corresponding coherent-state representations of the transition amplitude and
the generating functional in the zero-temperature quantum theory [9-11].
Particularly, when the functional integral is of Gaussian type, the
partition function and the generating functional can exactly be calculated
by means of the stationary-phase method without any uncertainty [9-11]. For
the case of interacting systems, the partition function and
finite-temperature Green's functions can be conveniently calculated from the
generating functional by the perturbation method.

The aim of this paper is to formulate the quantization of thermal fields in
the coherent-state representation. To be definite, we will choose the
thermal quantum electrodynamics (QED) and the thermal $\varphi ^4$ theory
[5,6] as examples to describe the quantization of scalar fields, fermion
fields and gauge fields. In this quantization, we will first derive correct
path-integral expressions of the partition functions and the generating
functionals of Green's functions for these fields. Then, we focus our
attention to the perturbation method of calculating the partition functions
and the generating functionals. In the zero-order approximation, the
partition functions and the generating functionals will be exactly
calculated by means of the stationary-phase method, giving results as the
same as those given previously from the theories formulated in the position
space.

The remainder of this paper is arranged as follows. In Sect. 2, we quote the
main results given in our previous paper for quantum statistical mechanics.
These results may straightforwardly be extended to the quantum field theory.
In Sect. 3, we describe the coherent-state representation of Hamiltonians
and actions for the thermal QED and $\varphi ^4$ theory which are needed for
quantizing the theories in the coherent-state representation. In Sect. 4,
the quantizations of the thermal QED and $\varphi ^4$ theory are
respectively performed in the coherent-state representation. In doing this,
we write out explicitly the generating functionals of thermal Green's
functions for the theories mentioned above. In particular, we pay our main
attention to deriving the perturbative expansions of the generating
functionals in the coherent-state representation. In Sect. 5, the
perturbative expansions of the generating functionals given in Sect.4 will
be transformed to the corresponding ones represented in the position space.
In the last section, some concluding remarks will be made.

\section{Path integral formulation of quantum statistics in the
coherent-state representation}

First, we start from the partition function for a grand canonical ensemble
which usually is written in the form [4-6] 
\begin{equation}
Z=Tre^{-\beta \widehat{K}}  \eqnum{1}
\end{equation}
where $\beta =\frac 1{kT}$ with $k$ and $T$ being the Boltzmann constant and
the temperature and 
\begin{equation}
\widehat{K}=\widehat{H}-\mu \widehat{N}  \eqnum{2}
\end{equation}
here $\mu $ is the chemical potential, $\widehat{H}$ and $\widehat{N}$ are
the Hamiltonian and particle-number operators respectively. In the
coherent-state representation, the trace in Eq. (1) will be represented by
an integral over the coherent states. To determine the concrete form of the
integral, it is convenient to start from an one-dimensional system. Its
partition function given in the particle-number representation is 
\begin{equation}
Z=\sum_{n=0}^\infty \left\langle n\right| e^{-\beta \widehat{K}}\left|
n\right\rangle .  \eqnum{3}
\end{equation}
Then, we use the completeness relation of the coherent states [3-10] 
\begin{equation}
\int D(a^{*}a)\mid a><a^{*}\mid =1  \eqnum{4}
\end{equation}
where $\mid a>$ denotes a normalized coherent state, i.e., the eigenstate of
the annihilation operator $\hat a$ with a complex eigenvalue $a$ [9-12] 
\begin{equation}
\widehat{a}\left| a\right\rangle =a\left| a\right\rangle   \eqnum{5}
\end{equation}
whose Hermitian conjugate is 
\begin{equation}
\left\langle a^{*}\right| \widehat{a}^{+}=a^{*}\left\langle a^{*}\right|  
\eqnum{6}
\end{equation}
and $D(a^{*}a)$ symbolizes the integration measure defined by 
\begin{equation}
D(a^{*}a)=\{
\begin{array}{cc}
\frac 1\pi da^{*}da, & \text{for bosons;} \\ 
da^{*}da, & \text{for fermions.}
\end{array}
\eqnum{7}
\end{equation}
In the above, we have used the eigenvalues $a$ and $a^{*}$ to designate the
eigenstates $\left| a\right\rangle $ and $\left\langle a^{*}\right| $,
respectively. Inserting Eq. (4) into Eq. (3), we have 
\begin{equation}
Z=\sum_{n=0}^\infty \int D(a^{*}a)D(a^{\prime *}a^{\prime })\left\langle
n\mid a^{\prime }\right\rangle \left\langle a^{\prime *}\right| e^{-\beta 
\widehat{K}}\left| a\right\rangle \left\langle a^{*}\mid n\right\rangle  
\eqnum{8}
\end{equation}
where 
\begin{equation}
\begin{array}{c}
\left\langle a^{*}\mid n\right\rangle =\frac 1{\sqrt{n!}}%
(a^{*})^ne^{-a^{*}a}, \\ 
\left\langle n\mid a^{\prime }\right\rangle =\frac 1{\sqrt{n!}}(a^{\prime
})^ne^{-a^{\prime *}a^{\prime }}
\end{array}
\eqnum{9}
\end{equation}
are the energy eigenfunctions given in the coherent-state representation
(Note: for fermions, $n=0,1$) [4-10]. The both eigenfunctions commute with
the matrix element $\left\langle a^{\prime *}\mid e^{-\beta \widehat{K}}\mid
a\right\rangle $ because the operator $\widehat{K}(\widehat{a}^{+},\widehat{a%
})$ generally is a polynomial of the operator $\widehat{a}^{+}\widehat{a}$
for fermion systems. In view of the expressions in Eq. (9) and the
commutation relation [3-10] 
\begin{equation}
a^{*}a^{\prime }=\pm a^{\prime }a^{*}  \eqnum{10}
\end{equation}
where the signs ''$+$'' and ''$-$'' are attributed to bosons and fermions
respectively, it is easy to see 
\begin{equation}
\left\langle n\mid a^{\prime }\right\rangle \left\langle a^{*}\mid
n\right\rangle =\left\langle \pm a^{*}\mid n\right\rangle \left\langle n\mid
a^{\prime }\right\rangle .  \eqnum{11}
\end{equation}
Substituting Eq. (11) in Eq. (8) and applying the completeness relations for
the particle-number and coherent states, one may find 
\begin{equation}
Z=\int D(a^{*}a)\left\langle \pm a^{*}\right| e^{-\beta \widehat{K}}\left|
a\right\rangle   \eqnum{12}
\end{equation}
where the plus and minus signs in front of $a^{*}$ belong to bosons and
fermions respectively.

To evaluate the matrix element in Eq. (12), we may, as usual, divide the
''time'' interval $\left[ 0,\beta \right] $ into $n$ equal and infinitesimal
parts, $\beta =n\varepsilon $. and then insert a completeness relation shown
in Eq. (4) at each dividing point. In this way, Eq. (12) may be represented
as [1, 3-6] 
\begin{equation}
\begin{array}{c}
Z=\int D(a^{*}a)\prod\limits_{i=1}^{n-1}D(a_i^{*}a_i)\left\langle \pm
a^{*}\right| e^{-\varepsilon \widehat{K}}\left| a_{n-1}\right\rangle \\ 
\times \left\langle a_{n-1}^{*}\right| e^{-\varepsilon \widehat{K}}\left|
a_{n-2}\right\rangle \cdots \cdots \left\langle a_1^{*}\right|
e^{-\varepsilon \widehat{K}}\left| a\right\rangle .
\end{array}
\eqnum{13}
\end{equation}
Since $\varepsilon $ is infinitesimal, we may write 
\begin{equation}
e^{-\varepsilon \widehat{K}(\widehat{a}^{+},\widehat{a})}=1-\varepsilon 
\widehat{K}(\widehat{a}^{+},\widehat{a})  \eqnum{14}
\end{equation}
where $\widehat{K}(\widehat{a}^{+},\widehat{a})$ is assumed to be normally
ordered. Noticing this fact, when applying the equations (5) and (6) and the
inner product of two coherent states [1, 3-12] 
\begin{equation}
\left\langle a_i^{*}\mid a_{i-1}\right\rangle =e^{-\frac 12a_i^{*}a_i-\frac 1%
2a_{i-1}^{*}a_{i-1}+a_i^{*}a_{i-1}}  \eqnum{15}
\end{equation}
which suits to the both of bosons and fermions, one can get from Eq. (13)
that 
\begin{equation}
\begin{array}{c}
Z=\int D(a^{*}a)e^{-a^{*}a}\int \prod\limits_{i=1}^{n-1}D(a_i^{*}a_i)\exp
\{-\varepsilon \sum\limits_{i=1}^nK(a_i^{*},a_{i-1}) \\ 
+\sum\limits_{i=1}^na_i^{*}a_{i-1}-\sum\limits_{i=1}^{n-1}a_i^{*}a_i\}
\end{array}
\eqnum{16}
\end{equation}
where we have set 
\begin{equation}
\pm a^{*}=a_n^{*}\text{ , }a=a_0.  \eqnum{17}
\end{equation}
It is noted that the factor $e^{-a^{*}a}$ in the first integrand comes from
the matrix elements $\left\langle \pm a^{*}\right| a_{n-1}\rangle $ and $%
\left\langle a_1^{*}\right| a\rangle $ and the last sum in the above
exponent is obtained by summing up the common terms $-\frac 12a_i^{*}a_i$
and $-\frac 12a_{i-1}^{*}a_{i-1}$ appearing in the exponents of the matrix
element $\langle a_i^{*}\mid a_{i-1}\rangle $ and its adjacent ones $\langle
a_{i+1}^{*}\mid a_i\rangle $ and $\langle a_{i-1}^{*}\mid a_{i-2}\rangle $.
As will be seen in Eq. (21), such a summation is essential to give a correct
time-dependence of the functional integrand in the partition function. The
last two sums in Eq. (16) can be rewritten in the form 
\begin{equation}
\begin{array}{c}
\sum\limits_{i=1}^na_i^{*}a_{i-1}-\sum\limits_{i=1}^{n-1}a_i^{*}a_i \\ 
=\frac 12a_n^{*}a_{n-1}+\frac 12a_1^{*}a_0+\frac \varepsilon 2%
\sum\limits_{i=1}^{n-1}[(\frac{a_{i+1}^{*}-a_i^{*}}\varepsilon )a_i-a_i^{*}(%
\frac{a_i-a_{i-1}}\varepsilon )].
\end{array}
\eqnum{18}
\end{equation}
Upon substituting Eq. (18) in Eq. (16) and taking the limit $\varepsilon
\rightarrow 0$, we obtain the path-integral expression of the partition
functions as follows: 
\begin{equation}
Z=\int D(a^{*}a)e^{-a^{*}a}\int {\frak D}(a^{*}a)e^{I(a^{*},a)}  \eqnum{19}
\end{equation}
where 
\begin{equation}
{\frak D}(a^{*}a)=\{ 
\begin{array}{cc}
\prod\limits_\tau \frac 1\pi da^{*}(\tau )da(\tau ), & \text{for bosons;} \\ 
\prod\limits_\tau da^{*}(\tau )da(\tau ), & \text{for fermions}
\end{array}
\eqnum{20}
\end{equation}
and 
\begin{equation}
\begin{array}{c}
I(a^{*},a)=\frac 12a^{*}(\beta )a(\beta )+\frac 12a^{*}(0)a(0)-\int_0^\beta
d\tau [\frac 12a^{*}(\tau )\dot a(\tau ) \\ 
-\frac 12\dot a^{*}(\tau )a(\tau )+K(a^{*}(\tau ),a(\tau ))] \\ 
=a^{*}(\beta )a(\beta )-\int_0^\beta d\tau [a^{*}(\tau )\dot a(\tau
)+K(a^{*}(\tau ),a(\tau ))
\end{array}
\eqnum{21}
\end{equation}
where the last equality is obtained from the first one by a partial
integration. In accordance with the definition given in Eq. (17), we see,
the path-integral is subject to the following boundary conditions 
\begin{equation}
a^{*}(\beta )=\pm a^{*},a(0)=a  \eqnum{22}
\end{equation}
where the signs ''$+$'' and ''$-$'' are written respectively for bosons and
fermions. Here it is noted that Eq. (22) does not implies $a(\beta )=\pm a$
and $a^{*}(0)=a^{*}$. Actually, we have no such boundary conditions.

For the systems with many degrees of freedom, the functional-integral
representation of the partition functions may directly be written out from
the results given in Eqs. (19) -(22) as long as the eigenvalues $a$ and $%
a^{*}$ are understood as column matrices $a=(a_1,a_2,\cdots ,a_k,\cdots )$
and $a^{*}=(a_1^{*},a_2^{*},\cdots ,a_k^{*},\cdots )$. Written explicitly,
we have 
\begin{equation}
Z=\int D(a^{*}a)e^{-a_k^{*}a_k}\int {\frak D}(a^{*}a)e^{I(a^{*},a)} 
\eqnum{23}
\end{equation}
where 
\begin{equation}
D(a^{*}a)=\{ 
\begin{array}{cc}
\prod\limits_k\frac 1\pi da_k^{*}da_k & \text{, for bosons;} \\ 
\prod\limits_kda_k^{*}da_k & \text{, for fermions,}
\end{array}
\eqnum{24}
\end{equation}
\begin{equation}
{\frak D}(a^{*}a)=\{ 
\begin{array}{cc}
\prod\limits_{k\tau }\frac 1\pi da_k^{*}(\tau )da_k(\tau ) & \text{, for
bosons;} \\ 
\prod\limits_{k\tau }da_k^{*}(\tau )da_k(\tau ) & \text{, for fermions}
\end{array}
\eqnum{25}
\end{equation}
and 
\begin{equation}
I(a^{*},a)=a_k^{*}(\beta )a_k(\beta )-\int_0^\beta d\tau [a_k^{*}(\tau )\dot 
a_k(\tau )+K(a_k^{*}(\tau ),a_k(\tau ))].  \eqnum{26}
\end{equation}
The boundary conditions in Eq. (22) now become 
\begin{equation}
a_k^{*}(\beta )=\pm a_k^{*}\text{ , }a_k(0)=a_k.  \eqnum{27}
\end{equation}
In Eqs. (23) and (26), the repeated indices imply the summations over $k$.
If the $k$ stands for a continuous index as in the case of quantum field
theory, the summations will be replaced by integrations over $k$.

It should be pointed out that in the previous derivation of the
coherent-state representation of the partition functions, the authors did
not use the expressions given in Eqs. (16) and (18). Instead, the matrix
element in Eq. (15) was directly chosen to be the starting point and recast
in the form [1-6]

\begin{equation}
\langle a_i^{*}\mid a_{i-1}\rangle =\exp \{-\frac \varepsilon 2[a_i^{*}(%
\frac{a_i-a_{i-1}}\varepsilon )-(\frac{a_i^{*}-a_{i-1}^{*}}\varepsilon
)a_{i-1}]\}.  \eqnum{28}
\end{equation}
Substituting the above expression into Eq. (13) and taking the limit $%
\varepsilon \rightarrow 0$, it follows [1-6]

\begin{equation}
Z=\int {\frak D}(a^{*}a)\exp \{-\int_0^\beta d\tau [\frac 12a^{*}(\tau )\dot 
a(\tau )-\frac 12\dot a^{*}(\tau )a(\tau )+K(a^{*}(\tau ),a(\tau ))]\}. 
\eqnum{29}
\end{equation}
Clearly, in the above derivation, the common terms appearing in the
exponents of adjacent matrix elements were not combined together. As a
result, the time-dependence of the integrand in Eq. (29) could not be given
correctly. In comparison with the previous result shown in Eq. (29), the
expression written in Eqs. (19)-(21) has two functional integrals. The first
integral which represents the trace in Eq. (1) is absent in Eq. (29). The
second integral is defined as the same as the integral in Eq. (29); but the
integrand are different from each other. In Eq. (19), there occur two
additional factors in the integrand : one is $e^{-a^{*}a}$ which comes from
the initial and final states in Eq. (13), another is $e^{\frac 12%
[a^{*}(\beta )a(\beta )+a^{*}(0)a(0)]}$ in which $a^{*}(\beta )$ and $a(0)$
are related to the boundary conditions shown in Eq. (22). These additional
factors are also absent in Eq. (29). As will be seen soon later, the
occurrence of these factors in the functional-integral expression is
essential to give correct calculated results.

To demonstrate the correctness of the expression given in Eqs. (23)-(27),
let us compute the partition function for the system whose Hamiltonian is of
harmonic oscillator-type as we meet in the cases of ideal gases and free
fields. In this case 
\begin{equation}
K(a^{*}a)=\omega _ka_k^{*}a_k  \eqnum{30}
\end{equation}
where $\omega _k=\varepsilon _k-\mu $ with $\varepsilon _k$ being the
particle energy and therefore Eq. (26) becomes 
\begin{equation}
I(a^{*},a)=a_k^{*}(\beta )a_k(\beta )-\int_0^\beta d\tau [a_k^{*}(\tau )\dot 
a_k(\tau )+\omega _ka_k^{*}(\tau )a_k(\tau )].  \eqnum{31}
\end{equation}
By the stationary-phase method which is established based on the property of
the Gaussian integral that the integral is equal to the extremum of its
integrand [8-11], we may write 
\begin{equation}
\int {\frak D}(a^{*}a)e^{I(a^{*},a)}=e^{I_0(a^{*},a)}  \eqnum{32}
\end{equation}
where $I_0(a^{*},a)$ is obtained from $I(a^{*},a)$ by replacing the
variables $a_k^{*}(\tau )$ and $a_k(\tau )$ in $I(a^{*},a)$ with those
values which are determined from the stationary condition $\delta
I(a^{*},a)=0$. From this condition and the boundary conditions in Eq. (27)
which implies $\delta a_k^{*}(\beta )=0$ and $\delta a_k(0)=0$, it is easy
to derive the following equations of motion [8-11] 
\begin{equation}
\dot a_k(\tau )+\omega _ka_k(\tau )=0,\text{ }\dot a_k^{*}(\tau )-\omega
_ka_k^{*}(\tau )=0.  \eqnum{33}
\end{equation}
Their solutions satisfying the boundary condition are 
\begin{equation}
a_k(\tau )=a_ke^{-\omega _k\tau }\text{ , }a_k^{*}(\tau )=\pm
a_k^{*}e^{\omega _k(\tau -\beta )}.  \eqnum{34}
\end{equation}
On substituting the above solutions into Eq. (31), we obtain 
\begin{equation}
I_0(a^{*},a)=\pm a_k^{*}a_ke^{-\omega _k\beta }.  \eqnum{35}
\end{equation}
With the functional integral given in Eqs. (32) and (35), the partition
functions in Eq. (23) become 
\begin{equation}
Z_0=\{ 
\begin{array}{cc}
\int D(a^{*}a)e^{-a_k^{*}a_k(1-e^{-\beta \omega _k})} & \text{ , for bosons;}
\\ 
\int D(a^{*}a)e^{-a_k^{*}a_k(1+e^{-\beta \omega _k})} & \text{ , for
fermions.}
\end{array}
\eqnum{36}
\end{equation}
For the boson case, the above integral can directly be calculated by
employing the integration formula [1]: 
\begin{equation}
\int D(a^{*}a)e^{-a^{*}(\lambda a-b)}f(a)=\frac 1\lambda f(\lambda ^{-1}b). 
\eqnum{37}
\end{equation}
The result is well-known, as shown in the following [3-6] 
\begin{equation}
Z_0=\prod\limits_k\frac 1{1-e^{-\beta \omega _k}}.  \eqnum{38}
\end{equation}
For the fermion case, by using the property of Grassmann algebra and the
integration formulas [3-10] 
\begin{equation}
\int da=\int da^{*}=0\text{ , }\int da^{*}a^{*}=\int daa=1,  \eqnum{39}
\end{equation}
one may easily compute the integral in Eq. (36) and gets the familiar result
[3-6] 
\begin{equation}
Z_0=\prod\limits_k(1+e^{-\beta \omega _k}).  \eqnum{40}
\end{equation}
It is noted that if the stationary-phase method is applied to the functional
integral in Eq. (29), one could not get the results as written in Eqs. (38)
and (40), showing the incorrectness of the previous functional-integral
representation for the partition functions.

Now let us turn to discuss the general case where the Hamiltonian can be
split into a free part and an interaction part. Correspondingly, we can
write 
\begin{equation}
K(a^{*},a)=K_0(a^{*},a)+H_I(a^{*},a)  \eqnum{41}
\end{equation}
where $K_0(a^{*},a)$ is the same as given in Eq. (30). In this case, to
evaluate the partition function, it is convenient to define a generating
functional through introducing external sources $j_k^{*}(\tau )$ and $%
j_k(\tau )$ such that [4,5] 
\begin{equation}
\begin{array}{c}
Z[j^{*},j]=\int D(a^{*}a)e^{-a_{^{*}k}a_k}\int {\frak D}(a^{*}a)e^{a_k^{*}(%
\beta )a_k(\beta )-\int_0^\beta d\tau [a_k^{*}(\tau )\stackrel{\bullet }{a}%
_k(\tau )+K(a^{*}a)-j_k^{*}(\tau )a_k(\tau )-a_k^{*}(\tau )j_k(\tau )]} \\ 
=e^{-\int_0^\beta d\tau H_I(\frac \delta {\delta j_k^{*}(\tau )},\pm \frac 
\delta {\delta j_k(\tau )})}Z_0[j^{*},j]
\end{array}
\eqnum{42}
\end{equation}
where the sings ''$+$'' and ''$-$'' in front of $\frac \delta {\delta
j_k(\tau )}$ refer to bosons and fermions respectively and $Z_0[j^{*},j]$ is
defined by 
\begin{equation}
Z_0[j^{*},j]=\int D(a^{*}a)e^{-a_k^{*}a_k}\int {\frak D}%
(a^{*}a)e^{I(a^{*},a;j^{*},j)}  \eqnum{43}
\end{equation}
in which 
\begin{equation}
\begin{array}{c}
I(a^{*},a;j^{*},j)=a_k^{*}(\beta )a_k(\beta )-\int_0^\beta d\tau
[a_k^{*}(\tau )\dot a_k(\tau ) \\ 
+\omega _ka_k^{*}(\tau )a_k(\tau )-j_k^{*}(\tau )a_k(\tau )-a_k^{*}(\tau
)j_k(\tau )].
\end{array}
\eqnum{44}
\end{equation}
Obviously, the integral in Eq. (43) is of Gaussian-type. Therefore, it can
be calculated by means of the stationary-phase method as will be shown in
detail in Sect. 4.

The exact partition functions can be obtained from the generating functional
in Eq. (42) by setting the external sources to be zero 
\begin{equation}
Z=Z[j^{*},j]\mid _{j^{*}=j=0.}  \eqnum{45}
\end{equation}
In particular, the generating functional is much useful to compute the
finite-temperature Green's functions. For simplicity, we take the two-point
Green's function as an example to show this point. In many-body theory, the
Green's function usually is defined in the operator formalism by [4,13] 
\begin{equation}
G_{kl}(\tau _1,\tau _2)=\frac 1ZTr\{e^{-\beta \widehat{K}}T[\widehat{a}%
_k(\tau _1)\widehat{a}_l^{+}(\tau _2)]\}=Tr\{e^{\beta (\Omega -\widehat{K}%
})T[\widehat{a}_k(\tau _1)\widehat{a}_l^{+}(\tau _2)]\}  \eqnum{46}
\end{equation}
where $0<\tau _1,\tau _2<\beta $, $\Omega =-\frac 1\beta \ln Z$ is the grand
canonical potential, $T$ denotes the ''time'' ordering operator, $\widehat{a}%
_k(\tau _1)$ and $\widehat{a}_l^{+}(\tau _2)$ represent the annihilation and
creation operators respectively. According to the procedure described in
Eqs. (12)-(22). it is clear to see that when taking $\tau _1$ and $\tau _2$
at two dividing points and applying the equations (5) and (6), the Green's
function may be expressed as a functional integral in the coherent-state
representation as follows: 
\begin{equation}
G_{kl}(\tau _1,\tau _2)=\frac 1Z\int D(a^{*}a)e^{-a_k^{*}a_k}\int {\frak D}%
(a^{*}a)a_k(\tau _1)a_l^{*}(\tau _2)e^{I(a^{*},a)}.  \eqnum{47}
\end{equation}
With the aid of the generating functional defined in Eq. (42), the above
Green's function may be represented as 
\begin{equation}
G_{kl}(\tau _1,\tau _2)=\pm \frac 1Z\frac{\delta ^2Z[j^{*},j]}{\delta
j_k^{*}(\tau _1)\delta j_l(\tau _2)}\mid _{j^{*}=j=0}  \eqnum{48}
\end{equation}
where the sings ''$+$'' and ''$-$'' belong to bosons and fermions
respectively.

\section{Coherent-state representation of thermal field Hamiltonians and
actions}

To write out explicitly a path-integral expression of a thermal field in the
coherent-state representation, we first need to formulate the field in the
coherent-state representation, namely, to give exact expressions of the
field Hamiltonian and action in the coherent-state representation. For this
purpose, we only need to work with the classical fields by using some
skilful treatments. In this section, we limit ourself to describe the
coherent-state representations of the thermal QED and $\varphi ^4$ theory .

Let us first start from the effective Lagrangian density of QED which
appears in the path-integral for the zero-temperature QED [9,10] 
\begin{equation}
{\cal L}=\bar \psi \{i\gamma ^\mu (\partial _\mu -ieA_\mu )-M\}\psi -\frac 14%
F^{\mu \nu }F_{\mu \nu }-\frac 1{2\alpha }(\partial ^\mu A_\mu )^2-\partial
^\mu \bar C\partial _\mu C  \eqnum{49}
\end{equation}
where $\psi $ and $\bar \psi $ represent the fermion field, $A_\mu $ is the
vector potential of photon field, $C$ and $\bar C$ designate the ghost field 
\begin{equation}
F_{\mu \nu }=\partial _\mu A_\nu -\partial _\nu A_\mu  \eqnum{50}
\end{equation}
and $M$ denotes the fermion mass. It is noted here that although there is no
coupling between the ghost field and the photon field, as will be shown
later, it is necessary to keep the ghost term in Eq. (49). For the sake of
simplicity, we will work in the Feynman gauge. In this gauge, the
Lagrangian, which is obtained from the above Lagrangian by applying the
Lorentz condition $\partial ^\mu A_\mu =0$, is of the form 
\begin{equation}
{\cal L}=\bar \psi \{i\gamma ^\mu (\partial _\mu -ieA_\mu )-M\}\psi -\frac 12%
\partial _\mu A_\nu \partial ^\mu A^\nu -\partial ^\mu \bar C\partial _\mu C
\eqnum{51}
\end{equation}
The above Lagrangian is written in the Minkowski metric where the $\gamma -$%
matrix is defined as $\gamma _0=\beta $ and $\vec \gamma =\beta \vec \alpha $
[10]. In the following, it is convenient to represent the Lagrangian in the
Euclidean metric with the imaginary time $\tau =it$ where $t$ is the real
time.

Since the path-integral in Eq. (42) is established in the first order (or
Hamiltonian) formalism, to perform the path-integral quantization of the
thermal QED in the coherent-state representation, we need to recast the
above Lagrangian in the first order form. In doing this, it is necessary to
introduce canonical conjugate momentum densities which are defined by
[9,10,14] 
\begin{equation}
\begin{array}{c}
\Pi _\psi =\frac{\partial {\cal L}}{\partial \partial _t\psi }=i\overline{%
\psi }\gamma ^0=i\psi ^{+}, \\ 
\Pi _{\overline{\psi }}=\frac{\partial {\cal L}}{\partial \partial _t%
\overline{\psi }}=0, \\ 
\Pi _\mu =\frac{\partial {\cal L}}{\partial \partial _tA^\mu }=-\partial
_tA_\mu , \\ 
\Pi =(\frac{\partial {\cal L}}{\partial \partial _tC})_R=-\partial _t%
\overline{C}, \\ 
\overline{\Pi }=(\frac{\partial {\cal L}}{\partial \partial _t\overline{C}}%
)_L=-\partial _tC
\end{array}
\eqnum{52}
\end{equation}
where the subscripts $R$ and $L$ mark the right and left-derivatives with
respect to the real time respectively. With the above momentum densities,
the Lagrangian in Eq. (51) can be represented as [9,10,14] 
\begin{equation}
{\cal L}=\Pi _\psi \partial _t\psi +\Pi ^\mu \partial _tA_\mu +\Pi \partial
_tC+\partial _t\overline{C}\overline{\Pi }-{\cal H}  \eqnum{53}
\end{equation}
where 
\begin{equation}
{\cal H}=\bar \psi (\overrightarrow{\gamma }\cdot \triangledown +m)\psi +%
\frac 12(\Pi _\mu )^2-\frac 12A_\mu \nabla ^2A_\mu -\Pi \overline{\Pi }+\bar 
C\nabla ^2C+ie\bar \psi \gamma _\mu \psi A_\mu  \eqnum{54}
\end{equation}
is the Hamiltonian density. This Hamiltonian density is now written in the
Euclidean metric for later convenience. The matrix $\gamma _\mu $ in this
metric is defined by $\gamma _4=\beta $ and $\vec \gamma =-i\beta \vec \alpha
$ [14]. It should be noted that the conjugate quantities $\Pi $ and $%
\overline{\Pi }$ for the ghost field are respectively defined by the
right-derivative and the left one as shown in Eq. (52) because only in this
way one can get correct results. This unusual definition originates from the
peculiar property of the ghost fields which iare scalar fields, but subject
to the commutation rule of Grassmann algebra.

In order to derive the coherent-state representation of the action of
thermal QED, one should employ the Fourier transformations for the canonical
variables of QED which are listed below. For the fermion field [5,6, 10], 
\begin{equation}
\begin{array}{c}
\psi (\vec x,\tau )=\int \frac{d^3p}{(2\pi )^{3/2}}[u^s(\vec p)b_s(\vec p%
,\tau )e^{i\vec p\cdot \vec x}+v^s(\vec p)d_s^{*}(\vec p,\tau )e^{-i\vec p%
\cdot \vec x}], \\ 
\overline{\psi }(\vec x,\tau )=\int \frac{d^3p}{(2\pi )^{3/2}}[\overline{u}%
^s(\vec p)b_s^{*}(\vec p,\tau )e^{-i\vec p\cdot \vec x}+\overline{v}^s(\vec p%
)d_s(\vec p,\tau )e^{i\vec p\cdot \vec x}]
\end{array}
\eqnum{55}
\end{equation}
where $u^s(\vec p)$ and $v^s(\vec p)$ are the spinor wave functions
satisfying the normalization conditions $u^{s+}(\vec p)u^s(\vec p)=v^{s+}(%
\vec p)v^s(\vec p)=1$, $b_s(\vec p,\tau )$ and $b_s^{*}(\vec p,\tau )$ are
the eigenvalues of the fermion annihilation and creation operators $\widehat{%
b}_s(\vec p,\tau )$ and $\widehat{b}_s^{+}(\vec p,\tau )$ which are defined
in the Heisenberg picture, $d_s(\vec p,\tau )$ and $d_s^{*}(\vec p,\tau )$
are the corresponding ones for antifermions. For the photon field [5,6,10], 
\begin{equation}
\begin{array}{c}
A_\mu (\vec x,\tau )=\int \frac{d^3k}{(2\pi )^{3/2}}\frac 1{\sqrt{2\omega (%
\vec k)}}\varepsilon _\mu ^\lambda (\vec k)[a_\lambda (\vec k,\tau )e^{i\vec 
k\cdot \vec x}+a_\lambda ^{*}(\vec k,\tau )e^{-i\vec k\cdot \vec x}], \\ 
\Pi _\mu (\vec x,\tau )=i\int \frac{d^3k}{(2\pi )^{3/2}}\sqrt{\frac{\omega (%
\vec k)}2}\epsilon _\mu ^\lambda (\vec k)[a_\lambda (\vec k,\tau )e^{i\vec k%
\cdot \vec x}-a_\lambda ^{*}(\vec k,\tau )e^{-i\vec k\cdot \vec x}]
\end{array}
\eqnum{56}
\end{equation}
where $\varepsilon _\mu ^\lambda (\overrightarrow{k})$ is the polarization
vector. The expression of $\Pi _\mu (\vec x,\tau )$ follows from the
definition in Eq. (52) and is consistent with the Fourier representation of
the free field. For the ghost fields which are always free for QED, we have 
\begin{equation}
\begin{array}{c}
\overline{C}(\vec x,\tau )=\int \frac{d^3q}{(2\pi )^{3/2}}\frac 1{\sqrt{%
2\omega (\vec q)}}[\overline{c}(\vec q,\tau )e^{i\vec q\cdot \vec x}+c^{*}(%
\vec q,\tau )e^{-i\vec q\cdot \vec x}], \\ 
C(\vec x,\tau )=\int \frac{d^3q}{(2\pi )^{3/2}}\frac 1{\sqrt{2\omega (\vec q)%
}}[c(\vec q,\tau )e^{i\vec q\cdot \vec x}+\overline{c}^{*}(\vec q,\tau )e^{-i%
\vec q\cdot \vec x}], \\ 
\Pi (\vec x,\tau )=i\int \frac{d^3q}{(2\pi )^{3/2}}\sqrt{\frac{\omega (\vec q%
)}2}[\overline{c}(\vec q,\tau )e^{i\vec q\cdot \vec x}-c^{*}(\vec q,\tau
)e^{-i\vec q\cdot \vec x}], \\ 
\overline{\Pi }(\vec x,\tau )=i\int \frac{d^3q}{(2\pi )^{3/2}}\sqrt{\frac{%
\omega (\vec q)}2}[c(\vec q,\tau )e^{i\vec q\cdot \vec x}-\overline{c}^{*}(%
\vec q,\tau )e^{-i\vec q\cdot \vec x}].
\end{array}
\eqnum{57}
\end{equation}
The expressions of $\Pi (\vec x,\tau )$ and $\overline{\Pi }(\vec x,\tau )$
also follow from the definitions in Eq. (52).

For simplifying the expressions of the Hamiltonian and action of the thermal
QED, it is convenient to use abbreviation notations. Define 
\begin{equation}
b_s^\theta (\vec p,\tau )= 
{b_s(\vec p,\tau ),\text{ }if\text{ }\theta =+, \atopwithdelims\{\} d_s^{*}(\vec p,\tau ),\text{ }if\text{ }\theta =-,}
\eqnum{58}
\end{equation}
\begin{equation}
W_s^\theta (\vec p)= 
{(2\pi )^{-3/2}u^s(\vec p),\text{ }if\text{ }\theta =+, \atopwithdelims\{\} (2\pi )^{-3/2}v^s(\vec p),\text{ }if\text{ }\theta =-}
\eqnum{59}
\end{equation}
and furthermore, set $\alpha =(\vec p,s,\theta )$ and 
\begin{equation}
\sum\limits_\alpha =\sum\limits_{s\theta }\int d^3p,  \eqnum{60}
\end{equation}
then , Eq. (55) may be represented as 
\begin{equation}
\begin{array}{c}
\psi (\vec x,\tau )=\sum\limits_\alpha W_\alpha b_\alpha (\tau )e^{i\theta 
\vec p\cdot \vec x}, \\ 
\overline{\psi }(\vec x,\tau )=\sum\limits_\alpha \overline{W}_\alpha
b_\alpha ^{*}(\tau )e^{-i\theta \vec p\cdot \vec x}.
\end{array}
\eqnum{61}
\end{equation}
Similarly, when we define 
\begin{equation}
a_{\lambda \theta }(\vec k,\tau )= 
{a_\lambda (\vec k,\tau ),\text{ }if\text{ }\theta =+, \atopwithdelims\{\} a_\lambda ^{*}(\vec k,\tau ),\text{ }if\text{ }\theta =-,}
\eqnum{62}
\end{equation}
\begin{equation}
\begin{array}{c}
A_{\mu \theta }^\lambda (\vec k)=(2\pi )^{-3/2}(2\omega (\vec k%
))^{-1/2}\epsilon _\mu ^\lambda (\vec k), \\ 
\Pi _{\mu \theta }^\lambda (\vec k)=i^\theta (2\pi )^{-3/2}[\omega (\vec q%
)/2]^{1/2}\epsilon _\mu ^\lambda (\vec k)
\end{array}
\eqnum{63}
\end{equation}
and furthermore, set $\alpha =(\vec k,\lambda ,\theta )$ and 
\begin{equation}
\sum\limits_\alpha =\sum\limits_{\lambda \theta }\int d^3k,  \eqnum{64}
\end{equation}
Eq. (56) can be written as 
\begin{equation}
\begin{array}{c}
A_\mu (\vec x,\tau )=\sum\limits_\alpha A_\mu ^\alpha a_\alpha (\tau
)e^{i\theta \vec k\cdot \vec x}, \\ 
\Pi _\mu (\vec x,\tau )=\sum\limits_\alpha \Pi _\mu ^\alpha a_\alpha (\tau
)e^{i\theta \vec k\cdot \vec x}.
\end{array}
\eqnum{65}
\end{equation}
For the ghost field, if we define 
\begin{equation}
c_\alpha ^\theta (\vec q,\tau )= 
{\overline{c}_a(\vec q,\tau ),\text{ }if\text{ }\theta =+, \atopwithdelims\{\} c_a^{*}(\vec q,\tau ),\text{ }if\text{ }\theta =-,}
\eqnum{66}
\end{equation}
\begin{equation}
\begin{array}{c}
G_\theta (\vec q)=(2\pi )^{-3/2}[2\omega (\vec q)]^{-1/2}, \\ 
\Pi _\theta (\vec q)=i^\theta (2\pi )^{-3/2}[\omega (\vec q)/2]^{1/2}
\end{array}
\eqnum{67}
\end{equation}
and furthermore set $\alpha =(\vec q,a,\theta )$ and 
\begin{equation}
\sum\limits_\alpha =\sum_{a\theta }\int d^3q  \eqnum{68}
\end{equation}
Eq. (57) will be expressed as 
\begin{equation}
\begin{array}{c}
\overline{C}^a(\vec x,\tau )=\sum\limits_\alpha G_\alpha c_\alpha (\tau
)e^{i\theta \vec q\cdot \vec x}, \\ 
C^a(\vec x,\tau )=\sum\limits_\alpha G_\alpha c_\alpha ^{*}(\tau
)e^{-i\theta \vec q\cdot \vec x}, \\ 
\Pi ^a(\vec x,\tau )=\sum\limits_\alpha \Pi _\alpha c_\alpha (\tau
)e^{i\theta \vec q\cdot \vec x}, \\ 
\overline{\Pi }^a(\vec x,\tau )=\sum\limits_\alpha \Pi _\alpha c_\alpha
^{*}(\tau )e^{-i\theta \vec q\cdot \vec x}.
\end{array}
\eqnum{69}
\end{equation}

Now we are ready to derive the expression of the action given by the
Lagrangian (53) in the coherent-state representation. For this purpose, it
is convenient to use the following boundary conditions of the fields [5,6]: 
\begin{equation}
\begin{array}{c}
\psi (\vec x,0)=\psi (\vec x),\text{ }\overline{\psi }(\vec x,0)=\overline{%
\psi }(\vec x), \\ 
\psi (\vec x,\beta )=-\psi (\vec x),\text{ }\overline{\psi }(\vec x,\beta )=-%
\overline{\psi }(\vec x),
\end{array}
\eqnum{70}
\end{equation}
\begin{equation}
\begin{array}{c}
A_\mu (\vec x,0)=A_\mu (\vec x,\beta )=A_\mu (\vec x), \\ 
\Pi _\mu (\vec x,0)=\Pi _\mu (\vec x,\beta )=\Pi _\mu (\vec x)
\end{array}
\eqnum{71}
\end{equation}
and 
\begin{equation}
\begin{array}{c}
\overline{C}(\vec x,0)=\overline{C}(\vec x,\beta )=\overline{C}(\vec x),%
\text{ }C(\vec x,0)=C(\vec x,\beta )=C(\vec x), \\ 
\overline{\Pi }(\vec x,0)=\overline{\Pi }(\vec x,\beta )=\overline{\Pi }(%
\vec x),\text{ }\Pi (\vec x,0)=\Pi (\vec x,\beta )=\Pi (\vec x).
\end{array}
\eqnum{72}
\end{equation}
Based on these boundary conditions, by partial integrations, the action may
be written in the form 
\begin{equation}
\begin{array}{c}
S=\int_0^\beta d\tau \int d^3x\{\frac 12[\psi ^{+}(\vec x,\tau )\dot \psi (%
\vec x,\tau )-\dot \psi ^{+}(\vec x,\tau )\psi (\vec x,\tau )] \\ 
+\frac i2[\Pi _\mu (\vec x,\tau )\dot A_\mu (\vec x,\tau )-\dot \Pi _\mu (%
\vec x,\tau )A_\mu (\vec x,\tau )] \\ 
+\frac i2[\Pi (\vec x,\tau )\dot C(\vec x,\tau )-\dot \Pi (\vec x,\tau )C(%
\vec x,\tau ) \\ 
+\overline{C}(\vec x,\tau )\stackrel{\cdot }{\overline{\Pi }}(\vec x,\tau )-%
\stackrel{\cdot }{\overline{C}}(\vec x,\tau )\overline{\Pi }(\vec x,\tau )]-%
{\cal H}(\vec x,\tau )\}
\end{array}
\eqnum{73}
\end{equation}
where the symbol $"\cdot "$ in $\dot \psi (\vec x,\tau ),\dot A_\mu (\vec x%
,\tau )\cdot \cdot \cdot $denotes the derivative of the fields with respect
to the imaginary time $\tau $. Upon substituting Eqs. (61), (65) and (69)
into Eq. (73), it is not difficult to get 
\begin{equation}
\begin{array}{c}
S=-\int_0^\beta d\tau \{\int d^3k\{\frac 12[b_s^{*}(\vec k,\tau )\dot b_s(%
\vec k,\tau )-\dot b_s^{*}(\vec k,\tau )b_s(\vec k,\tau )]+\frac 12[d_s^{*}(%
\vec k,\tau )\dot d_s(\vec k,\tau ) \\ 
-\dot d_s^{*}(\vec k,\tau )d_s(\vec k,\tau )]+\frac 12[a_\lambda ^{*}(\vec k%
,\tau )\dot a_\lambda (\vec k,\tau )-\dot a_\lambda ^{*}(\vec k,\tau
)a_\lambda (\vec k,\tau )]+\frac 12[\overline{c}^{*}(\vec k,\tau )\stackrel{%
\cdot }{\overline{c}}(\vec k,\tau ) \\ 
-\stackrel{\cdot }{\overline{c}^{*}}(\vec k,\tau )\overline{c}(\vec k,\tau
)-c^{*}(\vec k,\tau )\dot c(\vec k,\tau )+\dot c^{*}(\vec k,\tau )c(\vec k%
,\tau )]\}+H(\tau )\} \\ 
=-S_E
\end{array}
\eqnum{74}
\end{equation}
where $S_E$ is the action defined in the Euclidean metric and 
\begin{equation}
\begin{array}{c}
H(\tau )=\int d^3x{\cal H}(x)=H_0(\tau )+H_I(\tau )
\end{array}
\eqnum{75}
\end{equation}
in which $H_0(\tau )$ and $H_I(\tau )$ are respectively the free and
interaction Hamiltonians given in the coherent state representation. Their
expressions are 
\begin{equation}
\begin{array}{c}
H_0(\tau )=\sum\limits_\alpha \theta _\alpha \varepsilon _\alpha b_\alpha
^{*}(\tau )b_\alpha (\tau ) \\ 
+\frac 12\sum\limits_\alpha \omega _\alpha a_\alpha ^{*}(\tau )a_\alpha
(\tau )+\sum\limits_\alpha \omega _\alpha c_\alpha ^{*}(\tau )c_\alpha (\tau
)
\end{array}
\eqnum{76}
\end{equation}
and 
\begin{equation}
H_I(\tau )=\sum\limits_{\alpha \beta \gamma }f(\alpha \beta \gamma )b_\alpha
^{*}(\tau )b_\beta (\tau )a_\gamma (\tau )  \eqnum{77}
\end{equation}
where 
\begin{equation}
f(\alpha \beta \gamma )=ie(2\pi )^3\delta ^3(\theta _\alpha \vec p_\alpha
-\theta _\beta \vec p_\beta -\theta _\gamma \vec k_\gamma )\overline{W}%
_{s_\alpha }^{\theta _\alpha }(\vec p_\alpha )\gamma _\mu W_{s_\beta
}^{\theta _\beta }(\vec p_\beta )A_{\mu \lambda _\gamma }^{\theta _\gamma }(%
\vec k_\gamma ).  \eqnum{78}
\end{equation}
In the above, $\theta _\alpha \equiv \theta $, $\varepsilon _\alpha =(\vec p%
^2+M^2)^{1/2}$ is the fermion energy, $\omega _\alpha =\left| \vec k\right| $
is the energy for a photon or a ghost particle. It is emphasized that the
expressions in Eqs. (75)-(78) are just the Hamiltonian of QED appearing in
the path-integral shown in Eq. (42) where all the creation and annihilation
operators in the Hamiltonian which are written in a normal product are
replaced by their eigenvalues. It is noted that if one considers a grand
canonical ensemble of QED, the Hamiltonian in Eq. (74) should be replaced by 
$K(\tau )$ defined in Eq. (2). Employing the abbreviation notation as
denoted in Eqs. (58), (62) and (66) and letting $q_\alpha $ stand for $%
(a_\alpha ,b_\alpha ,c_\alpha )$, the action may compactly be represented as 
\begin{equation}
S_E=\int_0^\beta d\tau \{\sum\limits_\alpha \frac 12[q_\alpha ^{*}(\tau
)\circ \dot q_\alpha (\tau )-\dot q_\alpha ^{*}\circ q_\alpha (\tau
)]+H(\tau )\}  \eqnum{79}
\end{equation}
where we have defined 
\begin{equation}
q_\alpha ^{*}\circ q_\alpha =a_{\alpha ^{-}}a_{\alpha ^{+}}+b_\alpha
^{*}b_\alpha +\theta _\alpha c_\alpha ^{*}c_\alpha  \eqnum{80}
\end{equation}
and the Hamiltonian was represented in Eqs. (75)-(78). It is emphasized that
the $\theta _\alpha =\pm $ is now contained in the subscript $\alpha .$
Therefore, each $\alpha $ may takes $\alpha ^{+}$ and/or $\alpha ^{-}$ as
the first term in Eq. (80) does.

Now, let us turn to the $\varphi ^4$ theory whose Lagrangian at
zero-temperature is 
\begin{equation}
{\cal L}=\frac 12\partial ^\mu \varphi \partial _\mu \varphi -\frac 12%
m^2\varphi ^2-\frac 14\lambda \varphi ^4.  \eqnum{81}
\end{equation}
With the canonical conjugate momentum density defined by 
\begin{equation}
\Pi =\frac{\partial {\cal L}}{\partial \partial _t\varphi }=\partial
_t\varphi =i\dot \varphi ,  \eqnum{82}
\end{equation}
the Lagrangian in Eq. (81) at finite temperature may be written as 
\begin{equation}
{\cal L=}\pi \partial _t\varphi -{\cal H}  \eqnum{83}
\end{equation}
where 
\begin{equation}
{\cal H}=\frac 12\pi ^2+\frac 12(\nabla \varphi )^2+\frac 12m^2\varphi ^2+%
\frac 14\lambda \varphi ^4.  \eqnum{84}
\end{equation}
To derive the coherent-state representation of the action given by the
Lagrangian in Eq. (83), we need the following Fourier representation of the
canonical variables: 
\begin{equation}
\begin{array}{c}
\varphi (\vec x,\tau )=\int \frac{d^3k}{(2\pi )^{3/2}}\frac 1{\sqrt{2\omega (%
\vec k)}}[a(\vec k,\tau )e^{i\vec k\cdot \vec x}+a^{*}(\vec k,\tau )e^{-i%
\vec k\cdot \vec x}], \\ 
\pi (\vec x,\tau )=-i\int \frac{d^3k}{(2\pi )^{3/2}}\sqrt{\frac{\omega (\vec 
k)}2}[a(\vec k,\tau )e^{i\vec k\cdot \vec x}-a^{*}(\vec k,\tau )e^{-i\vec k%
\cdot \vec x}]
\end{array}
\eqnum{85}
\end{equation}
where $\omega (\vec k)=(\vec k^2+m^2)^{1/2}$. When we define 
\begin{equation}
a_\theta (\vec k,\tau )= 
{a(\vec k,\tau ),\text{ }if\text{ }\theta =+, \atopwithdelims\{\} a^{*}(\vec k,\tau ),\text{ }if\text{ }\theta =-,}
\eqnum{86}
\end{equation}
\begin{equation}
\begin{array}{c}
g_\theta (\vec k)=\frac 1{(2\pi )^{3/2}}\frac 1{\sqrt{2\omega (\vec k)}}, \\ 
\overline{g}_\theta (\vec k)=(-i)^\theta \frac 1{(2\pi )^{3/2}}\sqrt{\frac{%
\omega (\vec k)}2}
\end{array}
\eqnum{87}
\end{equation}
and furthermore set $\alpha =(\vec k,\theta )$ and 
\begin{equation}
\sum\limits_\alpha =\int d^3k\sum\limits_\theta ,  \eqnum{88}
\end{equation}
Eq. (85) can be written as 
\begin{equation}
\begin{array}{c}
\varphi (\vec x,\tau )=\sum\limits_\alpha g_\alpha a_\alpha (\tau
)e^{i\theta \vec k\cdot \vec x}, \\ 
\pi (\vec x,\tau )=\sum\limits_\alpha \overline{g}_\alpha a_\alpha (\tau
)e^{i\theta \vec k\cdot \vec x}.
\end{array}
\eqnum{89}
\end{equation}
With these expressions and considering the boundary conditions 
\begin{equation}
\begin{array}{c}
\varphi (\vec x,\beta )=\varphi (\vec x,0)=\varphi (\vec x), \\ 
\pi (\vec x,\beta )=\pi (\vec x,0)=\pi (\vec x),
\end{array}
\eqnum{90}
\end{equation}
the action given by the Lagrangian in Eq. (83) can be found to be 
\begin{equation}
\begin{array}{c}
S=\int_0^\beta d\tau \int d^3x\{\frac i2[\pi (\vec x,\tau )\dot \varphi (%
\vec x,\tau )-\dot \pi (\vec x,\tau )\varphi (\vec x,\tau )]-{\cal H}(\vec x%
,\tau )\} \\ 
=-\int_0^\beta d\tau \{\sum\limits_\alpha \frac 12[a_\alpha ^{*}(\tau )\dot a%
_\alpha (\tau )-\dot a_\alpha ^{*}(\tau )a_\alpha (\tau )]+H(\tau )\} \\ 
=-S_E
\end{array}
\eqnum{91}
\end{equation}
where 
\begin{equation}
H(\tau )=H_0(\tau )+H_I(\tau )  \eqnum{92}
\end{equation}
in which 
\begin{equation}
H_0(\tau )=\sum\limits_\alpha \omega _\alpha a_{\alpha ^{-}}(\tau )a_{\alpha
^{+}}(\tau )  \eqnum{93}
\end{equation}
is the free Hamiltonian and 
\begin{equation}
H_I(\tau )=\sum\limits_{\alpha \beta \gamma \delta }(2\pi )^3\delta
^3(\theta _\alpha \vec k_\alpha +\theta _\beta \vec k_\beta +\theta _\gamma 
\vec k_\gamma +\theta _\delta \vec k_\delta )\frac \lambda 4g_\alpha g_\beta
g_\gamma g_\delta a_\alpha a_\beta a_\gamma a_\delta  \eqnum{94}
\end{equation}
is the interaction Hamiltonian.

\section{Generating functional of Green's functions}

With the action $S_{E\text{ }}$given in the preceding section, the
quantization of the thermal fields in the coherent-state representation is
easily implemented by writing out its generating functional of thermal
Green's functions. According to the general formula shown in Eq. (42), the
QED generating functional can be formulated as 
\begin{equation}
\begin{array}{c}
Z[j]=\int D(q^{*}q)e^{-q^{*}\cdot q}\int {\frak D}(q^{*}q)\exp \{\frac 12%
[q^{*}(\beta )\cdot q(\beta ) \\ 
-q^{*}(0)\cdot q(0)]-S_E+\int_0^\beta d\tau j^{*}(\tau )\cdot q(\tau )\}
\end{array}
\eqnum{95}
\end{equation}
where we have defined 
\begin{equation}
q^{*}\cdot q=\frac 12a_\alpha ^{*}a_\alpha +\theta _\alpha b_\alpha
^{*}b_\alpha +c_\alpha ^{*}c_\alpha  \eqnum{96}
\end{equation}
and 
\begin{equation}
j^{*}\cdot q=\xi _\alpha ^{*}a_\alpha +\theta _\alpha (\eta _\alpha
^{*}b_\alpha +b_\alpha ^{*}\eta _\alpha +\zeta _\alpha ^{*}c_\alpha
+c_\alpha ^{*}\zeta _\alpha )  \eqnum{97}
\end{equation}
here $\xi _\alpha ,\eta _\alpha $ and $\zeta _\alpha $ are the sources for
photon, fermion and ghost particle respectively and the repeated index
implies summation. It is noted that the product $q^{*}\cdot q$ defined above
is different from the $q_\alpha ^{*}\circ q_\alpha $ defined in Eq. (80) in
the terms for fermion and ghost particle and the subscript $\alpha $ in Eqs.
(96) and (97) is also defined by containing $\theta _\alpha =\pm $ . In what
follows, we assign $\alpha ^{\pm }$ to represent the $\alpha $ with $\theta
_\alpha =\pm $. According to this notation, the sources in Eq. (97) are
specifically defined as follows: 
\begin{equation}
\begin{array}{c}
\xi _{\alpha ^{+}}=\xi _\alpha ,\text{ }\xi _{\alpha ^{-}}=\xi _\alpha ^{*},
\\ 
\eta _{\alpha ^{+}}=\eta _\alpha ,\text{ }\eta _{\alpha ^{-}}=\overline{\eta 
}_\alpha ^{*}, \\ 
\zeta _{\alpha ^{+}}=\zeta _\alpha ,\text{ }\zeta _{\alpha ^{-}}=\overline{%
\zeta }_\alpha ^{*}
\end{array}
\eqnum{98}
\end{equation}
where the subscript $\alpha $ on the right hand side of each equality no
longer contains $\theta _\alpha $ and the photon term in Eq. (96) $%
(1/2)a_\alpha ^{*}a_\alpha $ may be replaced by $a_{\alpha ^{-}}a_{\alpha
^{+}}$. The integration measures $D(q^{*}q)$ and ${\frak D}(q^{*}q)$ are
defined as shown in Eqs. (24) and (25).

Now we are interested in describing the perturbation method of calculating
the generating functional. Since the Hamiltonian can be split into two parts 
$H_0(\tau )$ and $H_I(\tau )$ as shown in Eqs. (76) and (77), the generating
functional in Eq. (95) may be perturbatively represented in the form 
\begin{equation}
Z[j]=\exp \{-\int_0^\beta d\tau H_I(\frac \delta {\delta j(\tau )})\}Z^0[j] 
\eqnum{99}
\end{equation}
where $Z^0[j]$ is the generating functional for the free system and the
exponential functional may be expanded in a Taylor series. In the above, the
commutativity between $H_I$ and $Z^0[j]$ has been considered. Obviously, the 
$Z^0[j]$ can be written as 
\begin{equation}
Z^0[j]=Z_p^0[\xi ]Z_f^0[\eta ]Z_g^0[\zeta ]  \eqnum{100}
\end{equation}
where $Z_p^0[\xi ]$, $Z_f^0[\eta ]$ and $Z_g^0[\zeta ]$ are the generating
functionals contributed from the free Hamiltonians of photons, fermions and
ghost particles respectively. They are separately and specifically described
below.

In view of the expressions in Eqs. (95), (74) and (76), the generating
functional $Z_p^0[\xi ]$ is of the form 
\begin{equation}
\begin{array}{c}
Z_p^0[\xi ]=\int D(a^{*}a)\exp \{-\int d^3k[a_\lambda ^{*}(\vec k)a_\lambda (%
\vec k)\} \\ 
\times \int {\frak D}(a^{*}a)\exp \{I_p(a_\lambda ^{*},a_\lambda ;\xi
_\lambda ^{*},\xi _\lambda )\}
\end{array}
\eqnum{101}
\end{equation}
where 
\begin{equation}
\begin{array}{c}
I_p(a_\lambda ^{*},a_\lambda ;\xi _\lambda ^{*},\xi _\lambda )=\int d^3k%
\frac 12[a_\lambda ^{*}(\vec k,\beta )a_\lambda (\vec k,\beta )+a_\lambda
^{*}(\vec k,0)a_\lambda (\vec k,0)] \\ 
-\int_0^\beta d\tau \int d^3k\{\frac 12[a_\lambda ^{*}(\vec k,\tau )\dot a%
_\lambda (\vec k,\tau )-\dot a_\lambda ^{*}(\vec k,\tau )a_\lambda (\vec k%
,\tau )]+\omega (\vec k)a_\lambda ^{*}(\vec k,\tau )a_\lambda (\vec k,\tau )
\\ 
-\xi _\lambda ^{*}(\vec k,\tau )a_\lambda (\vec k,\tau )-a_\lambda ^{*}(\vec 
k,\tau )\xi _\lambda (\vec k,\tau )\}
\end{array}
\eqnum{102}
\end{equation}
and 
\begin{equation}
\begin{array}{c}
D(a^{*}a)=\prod\limits_{\vec k\lambda }\frac 1\pi da_\lambda ^{*}(\vec k%
)da_\lambda (\vec k), \\ 
{\frak D}(a^{*}a)=\prod\limits_{\vec k\lambda \tau }\frac 1\pi da_\lambda
^{*}(\vec k,\tau )da_\lambda (\vec k,\tau ).
\end{array}
\eqnum{103}
\end{equation}
The subscript $\lambda $ in the above denotes the polarization. When we
perform a partial integration, Eq. (102) becomes 
\begin{equation}
\begin{array}{c}
I_p(a_\lambda ^{*},a_\lambda ;\xi _\lambda ^{*},\xi _\lambda )=\int
d^3ka_\lambda ^{*}(\vec k,\beta )a_\lambda (\vec k,\beta )-\int_0^\beta
d\tau \int d^3k\{a_\lambda ^{*}(\vec k,\tau )\dot a_\lambda (\vec k,\tau )
\\ 
+\omega (\vec k)a_\lambda ^{*}(\vec k,\tau )a_\lambda (\vec k,\tau )-\xi
_\lambda ^{*}(\vec k,\tau )a_\lambda (\vec k,\tau )-a_\lambda ^{*}(\vec k%
,\tau )\xi _\lambda (\vec k,\tau )\}.
\end{array}
\eqnum{104}
\end{equation}

For the generating functional $Z_f^0[\eta ]$, we can write 
\begin{equation}
\begin{array}{c}
Z_f^0[\eta ]=\int D(b^{*}bd^{*}d)\exp \{-\int d^3k[b_s^{*}(\vec k)b_s(\vec k%
)+d_s^{*}(\vec k)d_s(\vec k)]\} \\ 
\times \int {\frak D}(b^{*}bd^{*}d)\exp \{I_f(b_s^{*},b_s,d_s^{*},d_s;\eta
_s^{*},\eta _s,\overline{\eta }_s^{*},\overline{\eta }_s)\}
\end{array}
\eqnum{105}
\end{equation}
where 
\begin{equation}
\begin{array}{c}
I_f(b_s^{*},b_s,d_s^{*},d_s;\eta _s^{*},\eta _s,\overline{\eta }_s^{*},%
\overline{\eta }_s)=\int d^3k\frac 12[b_s^{*}(\vec k,\beta )b_s(\vec k,\beta
)+d_s^{*}(\vec k,\beta )d_s(\vec k,\beta ) \\ 
+b_s^{*}(\vec k,0)b_s(\vec k,0)+d_s^{*}(\vec k,0)d_s(\vec k,0)]-\int_0^\beta
d\tau \int d^3k\{\frac 12[b_s^{*}(\vec k,\tau )\dot b_s(\vec k,\tau ) \\ 
-\dot b_s^{*}(\vec k,\tau )b_s(\vec k,\tau )]+\frac 12[d_s^{*}(\vec k,\tau )%
\dot d_s(\vec k,\tau )-\dot d_s^{*}(\vec k,\tau )d_s(\vec k,\tau )] \\ 
+\varepsilon (\vec k)[b_s^{*}(\vec k,\tau )b_s(\vec k,\tau )+d_s^{*}(\vec k%
,\tau )d_s(\vec k,\tau )]-[\eta _s^{*}(\vec k,\tau )b_s(\vec k,\tau ) \\ 
+b_s^{*}(\vec k,\tau )\eta _s(\vec k,\tau )+\overline{\eta }_s^{*}(\vec k%
,\tau )d_s(\vec k,\tau )+d_s^{*}(\vec k,\tau )\overline{\eta }_s\vec k,\tau
)]\}
\end{array}
\eqnum{106}
\end{equation}
and 
\begin{equation}
\begin{array}{c}
D(b^{*}bd^{*}d)=\prod\limits_{\vec ks}db_s^{*}(\vec k)db_s(\vec k)dd_s^{*}(%
\vec k)dd_s(\vec k), \\ 
{\frak D}(b^{*}bd^{*}d)=\prod\limits_{\vec ks\tau }db_s^{*}(\vec k,\tau
)db_s(\vec k,\tau )dd_s^{*}(\vec k,\tau )dd_s(\vec k,\tau )
\end{array}
\eqnum{107}
\end{equation}
in which the subscript $s$ stands for the fermion spin. By a partial
integration over $\tau $, Eq. (106) may be given a simpler expression

\begin{equation}
\begin{array}{c}
I_f(b_s^{*},b_s,d_s^{*},d_s;\eta _s^{*},\eta _s,\overline{\eta }_s^{*},%
\overline{\eta }_s)=\int d^3k[b_s^{*}(\vec k,\beta )b_s(\vec k,\beta
)+d_s^{*}(\vec k,\beta )d_s(\vec k,\beta )] \\ 
-\int_0^\beta d\tau \int d^3k\{b_s^{*}(\vec k,\tau )\dot b_s(\vec k,\tau
)+d_s^{*}(\vec k,\tau )\dot d_s(\vec k,\tau ) \\ 
+\varepsilon (\vec k)[b_s^{*}(\vec k,\tau )b_s(\vec k,\tau )+d_s^{*}(\vec k%
,\tau )d_s(\vec k,\tau )]-[\eta _s^{*}(\vec k,\tau )b_s(\vec k,\tau ) \\ 
+b_s^{*}(\vec k,\tau )\eta _s(\vec k,\tau )+\overline{\eta }_s^{*}(\vec k%
,\tau )d_s(\vec k,\tau )+d_s^{*}(\vec k,\tau )\overline{\eta }_s\vec k,\tau
)]\}.
\end{array}
\eqnum{108}
\end{equation}

As for the generating functional $Z_g^0[\zeta ]$, we have 
\begin{equation}
\begin{array}{c}
Z_g^0[\zeta ]=\int D(\overline{c}^{*}\overline{c}cc^{*})\exp \{-\int d^3k[%
\overline{c}^{*}(\vec k)\overline{c}(\vec k)-c^{*}(\vec k)c(\vec k)]\} \\ 
\times \int {\frak D}(\overline{c}^{*}\overline{c}cc^{*})\exp \{I_g(c^{*},c,%
\overline{c}^{*},\overline{c};\zeta ^{*},\zeta ,\overline{\zeta }^{*},%
\overline{\zeta })\}
\end{array}
\eqnum{109}
\end{equation}
where 
\begin{equation}
\begin{array}{c}
I_g(c^{*},c,\overline{c}^{*},\overline{c};\zeta ^{*},\zeta ,\overline{\zeta }%
^{*},\overline{\zeta })=\int d^3k\frac 12[\overline{c}^{*}(\vec k,\beta )%
\overline{c}(\vec k,\beta )-c^{*}(\vec k,\beta )c(\vec k,\beta ) \\ 
+\overline{c}^{*}(\vec k,0)\overline{c}(\vec k,0)-c^{*}(\vec k,0)c(\vec k%
,0)]-\int_0^\beta d\tau \int d^3k\{\frac 12[\overline{c}^{*}(\vec k,\tau )%
\stackrel{\cdot }{\overline{c}}(\vec k,\tau ) \\ 
-\stackrel{\cdot }{\overline{c}^{*}}(\vec k,\tau )\overline{c}(\vec k,\tau
)]-\frac 12[c^{*}(\vec k,\tau )\dot c(\vec k,\tau )-\dot c^{*}(\vec k,\tau
)c(\vec k,\tau )] \\ 
+\omega (\vec k)[\overline{c}^{*}(\vec k,\tau )\overline{c}(\vec k,\tau
)-c^{*}(\vec k,\tau )c(\vec k,\tau )]-[\zeta ^{*}(\vec k,\tau )c(\vec k,\tau
) \\ 
+c^{*}(\vec k,\tau )\zeta (\vec k,\tau )+\overline{\zeta }^{*}(\vec k,\tau )%
\overline{c}(\vec k,\tau )+\overline{c}^{*}(\vec k,\tau )\overline{\zeta }(%
\vec k,\tau )]\}
\end{array}
\eqnum{110}
\end{equation}
and 
\begin{equation}
\begin{array}{c}
D(\overline{c}^{*}\overline{c}cc^{*})=\prod\limits_{\vec k}d\overline{c}^{*}(%
\vec k)d\overline{c}(\vec k)dc(\vec k)dc^{*}(\vec k), \\ 
{\frak D}(\overline{c}^{*}\overline{c}cc^{*})=\prod\limits_{\vec k\tau }d%
\overline{c}^{*}(\vec k,\tau )d\overline{c}(\vec k,\tau )dc(\vec k,\tau
)dc^{*}(\vec k,\tau ).
\end{array}
\eqnum{111}
\end{equation}
After a partial integration, Eq. (110) is reduced to 
\begin{equation}
\begin{array}{c}
I_g(c^{*},c,\overline{c}^{*},\overline{c};\zeta ^{*},\zeta ,\overline{\zeta }%
^{*},\overline{\zeta })=\int d^3k[\overline{c}^{*}(\vec k,\beta )\overline{c}%
(\vec k,\beta )-c^{*}(\vec k,\beta )c(\vec k,\beta ) \\ 
-\int_0^\beta d\tau \int d^3k\{\overline{c}^{*}(\vec k,\tau )\stackrel{\cdot 
}{\overline{c}}(\vec k,\tau )-c^{*}(\vec k,\tau )\dot c(\vec k,\tau )+\omega
(\vec k)[\overline{c}^{*}(\vec k,\tau )\overline{c}(\vec k,\tau ) \\ 
-c^{*}(\vec k,\tau )c(\vec k,\tau )]-[\zeta ^{*}(\vec k,\tau )c(\vec k,\tau
)+c^{*}(\vec k,\tau )\zeta (\vec k,\tau )+\overline{\zeta }^{*}(\vec k,\tau )%
\overline{c}(\vec k,\tau ) \\ 
+\overline{c}^{*}(\vec k,\tau )\overline{\zeta }(\vec k,\tau )]\}.
\end{array}
\eqnum{112}
\end{equation}
Here it is noted that all the terms related to the quantities $c^{*}$ and $c$
are opposite in sign to the terms related to the $\overline{c}^{*}$ and $%
\overline{c}$ and, correspondingly, the definitions of the integration
measures for these quantities, as shown in Eq. (111), are different from
each other in the order of the differentials.

The generating functionals in Eqs. (101), (105) and (109) are of
Gaussian-type, therefore, they can exactly be calculated by the
stationary-phase method. First, we calculate the functional integral $%
Z_p^0[\xi ]$. According to the stationary-phase method, the functional $%
Z_p^0[\xi ]$ can be represented in the form 
\begin{equation}
Z_p^0[\xi ]=\int D(a^{*}a)\exp \{-\int d^3k[a_\lambda ^{*}(\vec k)a_\lambda (%
\vec k)+I_p^0(a_\lambda ^{*},a_\lambda ;\xi _\lambda ^{*},\xi _\lambda )]\} 
\eqnum{113}
\end{equation}
where $I_p^0(a_\lambda ^{*},a_\lambda ;\xi _\lambda ^{*},\xi _\lambda )$ is
given by the stationary condition $\delta I_p(a_\lambda ^{*},a_\lambda ;\xi
_\lambda ^{*},\xi _\lambda )=0.$ By this condition and considering the
boundary conditions [4-6]: 
\begin{equation}
a_\lambda ^{*}(\vec k,\beta )=a_\lambda ^{*}(\vec k)\text{, }a_\lambda (\vec 
k,0)=a_\lambda (\vec k),  \eqnum{114}
\end{equation}
one may derive from Eq. (102) or (104) the following inhomogeneous equations
of motion [7,9]: 
\begin{equation}
\begin{array}{c}
\dot a_\lambda (\vec k,\tau )+\omega (\vec k)a_\lambda (\vec k,\tau )=\xi
_\lambda (\vec k,\tau ), \\ 
\dot a_\lambda ^{*}(\vec k,\tau )-\omega (\vec k)a_\lambda ^{*}(\vec k,\tau
)=-\xi _\lambda ^{*}(\vec k,\tau ).
\end{array}
\eqnum{115}
\end{equation}
In accordance with the general method of solving such a kind of equations,
one may first solve the homogeneous linear equations as written in Eq. (33).
Based on the solutions shown in Eq. (34) and the boundary condition denoted
in Eq. (114), one may assume [7,9] 
\begin{equation}
\begin{array}{c}
a_\lambda (\vec k,\tau )=[a_\lambda (\vec k)+u_\lambda (\vec k,\tau
)]e^{-\omega (\vec k)\tau }, \\ 
a_\lambda ^{*}(\vec k,\tau )=[a_\lambda ^{*}(\vec k)+u_\lambda ^{*}(\vec k%
,\tau )]e^{\omega (\vec k)(\tau -\beta )}
\end{array}
\eqnum{116}
\end{equation}
where the unknown functions $u_\lambda (\vec k,\tau )$ and $u_\lambda ^{*}(%
\vec k,\tau )$ are required to satisfy the boundary conditions [7,9]: 
\begin{equation}
u_\lambda (\vec k,0)=u_\lambda (\vec k,\beta )=u_\lambda ^{*}(\vec k%
,0)=u_\lambda ^{*}(\vec k,\beta )=0.  \eqnum{117}
\end{equation}
Inserting Eq. (116) into Eq. (115), we find 
\begin{equation}
\begin{array}{c}
\dot u_\lambda (\tau )=\xi _\lambda (\vec k,\tau )e^{\omega (\vec k)\tau },
\\ 
\dot u_\lambda ^{*}(\tau )=-\xi _\lambda ^{*}(\vec k,\tau )e^{\omega (\vec k%
)(\beta -\tau )}.
\end{array}
\eqnum{118}
\end{equation}
Integrating these two equations and applying the boundary conditions in Eq.
(117), one can get 
\begin{equation}
\begin{array}{c}
u_\lambda (\vec k,\tau )=\int_0^\tau d\tau ^{\prime }e^{\omega (\vec k)\tau
^{\prime }}\xi _\lambda (\vec k,\tau ^{\prime }), \\ 
u_\lambda ^{*}(\vec k,\tau )=-\int_\beta ^\tau d\tau ^{\prime }e^{\omega (%
\vec k)(\beta -\tau ^{\prime })}\xi _\lambda ^{*}(\vec k,\tau ^{\prime }).
\end{array}
\eqnum{119}
\end{equation}
Substitution of these solutions into Eq. (116) yields [7,9] 
\begin{equation}
\begin{array}{c}
a_\lambda (\vec k,\tau )=a_\lambda (\vec k)e^{-\omega (\vec k)\tau
}+\int_0^\tau d\tau ^{\prime }e^{-\omega (\vec k)(\tau -\tau ^{\prime })}\xi
_\lambda (\vec k,\tau ^{\prime }), \\ 
a_\lambda ^{*}(\vec k,\tau )=a_\lambda ^{*}(\vec k)e^{\omega (\vec k)(\tau
-\beta )}+\int_\tau ^\beta d\tau ^{\prime }e^{\omega (\vec k)(\tau -\tau
^{\prime })}\xi _\lambda ^{*}(\vec k,\tau ^{\prime }).
\end{array}
\eqnum{120}
\end{equation}
When Eq. (120) is inserted into Eq. (102) or Eq. (104), one may obtain the $%
I_p^0(a_\lambda ^{*},a_\lambda ;\xi _\lambda ^{*},\xi _\lambda )$ which
leads to an expression of Eq. (113) such that 
\begin{equation}
\begin{array}{c}
Z_p^0[\xi ]=\int D(a^{*}a)\exp \{-\int d^3k[a_\lambda ^{*}(\vec k)a_\lambda (%
\vec k)(1-e^{-\beta \omega (\vec k)})-a_\lambda ^{*}(\vec k)e^{-\beta \omega
(\vec k)} \\ 
\times \int_0^\beta d\tau e^{\omega (\vec k)\tau }\xi _\lambda (\vec k,\tau
)-\int_0^\beta d\tau e^{-\omega (\vec k)\tau }\xi _\lambda ^{*}(\vec k,\tau
)a_\lambda (\vec k)] \\ 
+\int_0^\beta d\tau _1\int_0^\beta d\tau _2\int d^3k\xi _\lambda ^{*}(\vec k%
,\tau _1)\theta (\tau _1-\tau _2)e^{-\omega (\vec k)(\tau _1-\tau _2)}\xi
_\lambda (\vec k,\tau _2)\}.
\end{array}
\eqnum{121}
\end{equation}
The above integral over $a_\lambda ^{*}(\vec k)$ and $a_\lambda (\vec k)$
can easily be calculated by applying the integration formula denoted in Eq.
(37) when we set 
\begin{equation}
\begin{array}{c}
\lambda =1-e^{-\beta \omega (\vec k)}, \\ 
b=e^{-\beta \omega (\vec k)}\int_0^\beta d\tau e^{\omega (\vec k)\tau }\xi
_\lambda (\vec k,\tau ), \\ 
f(a)=\int_0^\beta d\tau e^{-\omega (\vec k)\tau }\xi _\lambda ^{*}(\vec k%
,\tau )a_\lambda (\vec k).
\end{array}
\eqnum{122}
\end{equation}
The result is 
\begin{equation}
Z_p^0[\xi ]=Z_p^0\exp \{-\int_0^\beta d\tau _1\int_0^\beta d\tau _2\int
d^3k\xi ^{\lambda *}(\vec k,\tau _1)\Delta _{\lambda \lambda ^{\prime }}(%
\vec k,\tau _1-\tau _2)\xi ^{\lambda ^{\prime }}(\vec k,\tau _2)\} 
\eqnum{123}
\end{equation}
where 
\begin{equation}
Z_p^0=\prod\limits_{\overrightarrow{k}\lambda }[1-e^{-\beta \omega (\vec k%
)}]^{-1}=\prod\limits_{\overrightarrow{k}}[1-e^{-\beta \omega (\vec k)}]^{-4}
\eqnum{124}
\end{equation}
which is precisely the partition function contributed from the free photons
[5,6] and 
\begin{equation}
\Delta _{\lambda \lambda ^{\prime }}(\vec k,\tau _1-\tau _2)=g_{\lambda
\lambda ^{\prime }}\Delta _b(\vec k,\tau _1-\tau _2)  \eqnum{125}
\end{equation}
with

\begin{equation}
\Delta _b(\vec k,\tau _1-\tau _2)=[\theta (\tau _1-\tau _2)-(1-e^{\beta
\omega (\vec k)})^{-1}]e^{-\omega (\vec k)(\tau _1-\tau _2)}  \eqnum{126}
\end{equation}
is the free photon propagator given in the Feynman gauge and in the
Minkowski metric \{Note: in Euclidean metric, $g_{\lambda \lambda ^{\prime
}}\rightarrow -\delta _{\lambda \lambda ^{\prime }}$). When we interchange
the integration variables $\tau _1$ and $\tau _2$ and make a transformation $%
\vec k\rightarrow -\vec k$ in Eq. (123), by considering the relation 
\begin{equation}
\xi _\lambda ^{*}(\vec k,\tau )=\xi _\lambda (-\vec k,\tau )  \eqnum{127}
\end{equation}
which will be interpreted in the next section, one may find that the
propagator in Eq. (126) can be represented in the form 
\begin{equation}
\Delta _b(\vec k,\tau _1-\tau _2)=\frac 12[\overline{n}_b(\vec k)e^{-\omega (%
\vec k)\left| \tau _1-\tau _2\right| }-n_b(\vec k)e^{\omega (\vec k)\left|
\tau _1-\tau _2\right| }]  \eqnum{128}
\end{equation}
where 
\begin{equation}
\overline{n}_b(\vec k)=(1-e^{-\beta \varepsilon (\vec k)})^{-1},\text{ }n_b(%
\vec k)=(1-e^{\beta \varepsilon (\vec k)})^{-1}  \eqnum{129}
\end{equation}
which are just the boson distribution functions [3-6,13].

Let us turn to calculation of the functional integral in Eq. (105). On the
basis of stationary-phase method, we can write

\begin{equation}
\begin{array}{c}
Z_f^0[\eta ]=\int D(b^{*}bd^{*}d)\exp \{-\int d^3k[b_s^{*}(\vec k)b_s(\vec k%
)+d_s^{*}(\vec k)d_s(\vec k)]\} \\ 
\times \exp \{I_f^0(b_s^{*},b_s,d_s^{*},d_s;\eta _s^{*},\eta _s,\overline{%
\eta }_s^{*},\overline{\eta }_s)\}
\end{array}
\eqnum{130}
\end{equation}
where $I_f^0(b_s^{*},b_s,d_s^{*},d_s;\eta _s^{*},\eta _s,\overline{\eta }%
_s^{*},\overline{\eta }_s)$ will be obtained from Eq. (106) or Eq. (108) by
the stationary condition $\delta I_f(b_s^{*},b_s,d_s^{*},d_s;\eta
_s^{*},\eta _s,\overline{\eta }_s^{*},\overline{\eta }_s)=0$. From this
condition and the boundary conditions [4-6]

\begin{equation}
\begin{array}{c}
b_s^{*}(\vec k,\beta )=-b_s^{*}(\vec k),\text{ }b_s(\vec k,0)=b_s(\vec k),
\\ 
d_s^{*}(\vec k,\beta )=-d_s^{*}(\vec k),\text{ }d_s(\vec k,0)=d_s(\vec k),
\end{array}
\eqnum{131}
\end{equation}
one may deduce from Eq. (106) or Eq. (108) the following equations [7,9] 
\begin{equation}
\begin{array}{c}
\dot b_s(\vec k,\tau )+\varepsilon (\vec k)b_s(\vec k,\tau )=\eta _s(\vec k%
,\tau ), \\ 
\dot b_s^{*}(\vec k,\tau )-\varepsilon (\vec k)b_s^{*}(\vec k,\tau )=-\eta
_s^{*}(\vec k,\tau ), \\ 
\dot d_s(\vec k,\tau )+\varepsilon (\vec k)d_s(\vec k,\tau )=\overline{\eta }%
_s(\vec k,\tau ), \\ 
\dot d_s^{*}(\vec k,\tau )-\varepsilon (\vec k)d_s^{*}(\vec k,\tau )=-%
\overline{\eta }_s^{*}(\vec k,\tau ).
\end{array}
\eqnum{132}
\end{equation}
Following the procedure described in Eqs. (115)-(120), the solutions to the
above equations, which satisfies the boundary conditions in Eq. (131) and
the conditions like those in Eq. (117), can be found to be [7,9] 
\begin{equation}
\begin{array}{c}
b_s(\vec k,\tau )=b_s(\vec k)e^{-\varepsilon (\vec k)\tau }+\int_0^\tau
d\tau ^{\prime }e^{-\varepsilon (\vec k)(\tau -\tau ^{\prime })}\eta _s(\vec 
k,\tau ^{\prime }), \\ 
b_s^{*}(\vec k,\tau )=-b_s^{*}(\vec k)e^{\varepsilon (\vec k)(\tau -\beta
)}+\int_\tau ^\beta d\tau ^{\prime }e^{\varepsilon (\vec k)(\tau -\tau
^{\prime })}\eta _s^{*}(\vec k,\tau ^{\prime }), \\ 
d_s(\vec k,\tau )=d_s(\vec k)e^{-\varepsilon (\vec k)\tau }+\int_0^\tau
d\tau ^{\prime }e^{-\varepsilon (\vec k)(\tau -\tau ^{\prime })}\overline{%
\eta }_s(\vec k,\tau ^{\prime }), \\ 
d_s^{*}(\vec k,\tau )=-d_s^{*}(\vec k)e^{\varepsilon (\vec k)(\tau -\beta
)}+\int_\tau ^\beta d\tau ^{\prime }e^{\varepsilon (\vec k)(\tau -\tau
^{\prime })}\overline{\eta }_s^{*}(\vec k,\tau ^{\prime }).
\end{array}
\eqnum{133}
\end{equation}
Substituting the above solutions into Eq. (106) or (108), we find 
\begin{equation}
\begin{array}{c}
I_f^0(b_s^{*},b_s,d_s^{*},d_s;\eta _s^{*},\eta _s,\overline{\eta }_s^{*},%
\overline{\eta }_s)=\int d^3k\{-e^{-\beta \varepsilon (\vec k)}[b_s^{*}(\vec 
k)b_s(\vec k)+d_s^{*}(\vec k)d_s(\vec k)] \\ 
+\int_0^\beta d\tau e^{-\varepsilon (\vec k)\tau }[\eta _s^{*}(\vec k,\tau
)b_s(\vec k)+\overline{\eta }_s^{*}(\vec k,\tau )d_s(\vec k)]-e^{-\beta
\varepsilon (\vec k)}\int_0^\beta d\tau e^{\varepsilon (\vec k)\tau } \\ 
\times [b_s^{*}(\vec k)\eta _s(\vec k,\tau )+d_s^{*}(\vec k)\overline{\eta }%
_s(\vec k,\tau )]\}+B[\eta _s^{*},\eta _s,\overline{\eta }_s^{*},\overline{%
\eta }_s]
\end{array}
\eqnum{134}
\end{equation}
where 
\begin{equation}
\begin{array}{c}
B[\eta _s^{*},\eta _s,\overline{\eta }_s^{*},\overline{\eta }_s]=\frac 12%
\int_0^\beta d\tau _1\int_0^\beta d\tau _2\int d^3k\{\theta (\tau _1-\tau
_2)e^{-\varepsilon (\vec k)(\tau _1-\tau _2)} \\ 
\times [\eta _s^{*}(\vec k,\tau _1)\eta _s(\vec k,\tau _2)+\overline{\eta }%
_s^{*}(\vec k,\tau _1)\overline{\eta }_s(\vec k,\tau _2)]+\theta (\tau
_2-\tau _1)e^{-\varepsilon (\vec k)(\tau _2-\tau _1)} \\ 
\times [\eta _s^{*}(\vec k,\tau _2)\eta _s(\vec k,\tau _1)+\overline{\eta }%
_s^{*}(\vec k,\tau _2)\overline{\eta }_s(\vec k,\tau _1)]\}.
\end{array}
\eqnum{135}
\end{equation}
On inserting Eq. (134) into Eq. (130), we have 
\begin{equation}
Z_f^0[\eta ]=A[\eta _s^{*},\eta _s,\overline{\eta }_s^{*},\overline{\eta }%
_s]e^{B[\eta _s^{*},\eta _s,\overline{\eta }_s^{*},\overline{\eta }_s]} 
\eqnum{136}
\end{equation}
where 
\begin{equation}
\begin{array}{c}
A[\eta _s^{*},\eta _s,\overline{\eta }_s^{*},\overline{\eta }_s]=\int
D(b^{*}b)\exp \{-\int d^3k[b_s^{*}(\vec k)b_s(\vec k)(1+e^{-\beta
\varepsilon (\vec k)}) \\ 
+e^{-\beta \varepsilon (\overrightarrow{k})}b_s^{*}(\vec k)\int_0^\beta
d\tau e^{\varepsilon (\vec k)\tau }\eta _s(\vec k,\tau )-\int_0^\beta d\tau
e^{-\varepsilon (\vec k)\tau }\eta _s^{*}(\vec k,\tau )b_s(\vec k)]\} \\ 
\times \int D(d^{*}d)\exp \{-\int d^3k[d_s^{*}(\vec k)d_s(\vec k%
)(1+e^{-\beta \varepsilon (\vec k)})+e^{-\beta \varepsilon (\vec k)}d_s^{*}(%
\vec k) \\ 
\times \int_0^\beta d\tau e^{\varepsilon (\vec k)\tau }\overline{\eta }_s(%
\vec k,\tau )-\int_0^\beta d\tau e^{-\varepsilon (\vec k)\tau }\overline{%
\eta }_s^{*}(\vec k,\tau )d_s(\vec k)]\}
\end{array}
\eqnum{137}
\end{equation}
here the fact that the two integrals over \{$b^{*},b\}$ and \{$d^{*},d\}$
commute with each other has been considered. Obviously, each of the above
integrals can easily be calculated by applying the integration formulas
shown in Eq. (39). The result is 
\begin{equation}
\begin{array}{c}
A[\eta _s^{*},\eta _s,\overline{\eta }_s^{*},\overline{\eta }_s]=Z_f^0\exp
\{-\frac 12\int_0^\beta d\tau _1\int_0^\beta d\tau _2\int d^3k(1+e^{\beta
\varepsilon (\vec k)})^{-1} \\ 
\times \{e^{-\varepsilon (\vec k)(\tau _1-\tau _2)}[\eta _s^{*}(\vec k,\tau
_1)\eta _s(\vec k,\tau _2)+\overline{\eta }_s^{*}(\vec k,\tau _1)\overline{%
\eta }_s(\vec k,\tau _2)] \\ 
+e^{-\varepsilon (\vec k)(\tau _2-\tau _1)}[\eta _s^{*}(\vec k,\tau _2)\eta
_s(\vec k,\tau _1)+\overline{\eta }_s^{*}(\vec k,\tau _2)\overline{\eta }_s(%
\vec k,\tau _1)]\}\}
\end{array}
\eqnum{138}
\end{equation}
where

\begin{equation}
Z_f^0=\prod\limits_{\overrightarrow{k}s}[1+e^{-\beta \varepsilon (\vec k)}]^2
\eqnum{139}
\end{equation}
which just is the partition function contributed from free fermions and
antifermions [4-6,13]. It is noted that the two terms in the exponent are
equal to one another as seen from the interchange of the integration
variables $\tau _1$ and $\tau _2$. After Eqs. (135) and (138) are
substituted in Eq. (136), we get 
\begin{equation}
\begin{array}{c}
Z_f^0[\eta ]=Z_f^0\exp \{\int_0^\beta d\tau _1\int_0^\beta d\tau _2\int
d^3k\{[\theta (\tau _1-\tau _2)-(1+e^{\beta \varepsilon (\vec k%
)})^{-1}]e^{-\varepsilon (\vec k)(\tau _1-\tau _2)} \\ 
\times [\eta _s^{*}(\vec k,\tau _1)\eta _s(\vec k,\tau _2)+\overline{\eta }%
_s^{*}(\vec k,\tau _1)\overline{\eta }_s(\vec k,\tau _2)]+[\theta (\tau
_2-\tau _1)-(1+e^{\beta \varepsilon (\vec k)})^{-1}] \\ 
\times e^{-\varepsilon (\vec k)(\tau _2-\tau _1)}[\eta _s^{*}(\vec k,\tau
_2)\eta _s(\vec k,\tau _1)+\overline{\eta }_s^{*}(\vec k,\tau _2)\overline{%
\eta }_s(\vec k,\tau _1)]\}
\end{array}
\eqnum{140}
\end{equation}
When we interchange the variables $\tau _1$ and $\tau _2$, set $\vec k%
\rightarrow -\vec k$ in the second term of the above integrals and note the
relation 
\begin{equation}
\overline{\eta }_s^{*}(\vec k,\tau _2)\overline{\eta }_s(\vec k,\tau
_1)=\eta _s^{*}(-\vec k,\tau _1)\eta _s(-\vec k,\tau _2)  \eqnum{141}
\end{equation}
which will be proved in the next section, the functional $Z_f^0[\eta ]$ will
eventually be represented as 
\begin{equation}
\begin{array}{c}
Z_f^0[\eta ]=Z_f^0\exp \{\int_0^\beta d\tau _1\int_0^\beta d\tau _2\int
d^3k[\eta _s^{*}(\vec k,\tau _1)\Delta _f^{ss^{\prime }}(\vec k,\tau _1-\tau
_2)\eta _{s^{\prime }}(\vec k,\tau _2) \\ 
+\overline{\eta }_s^{*}(\vec k,\tau _1)\Delta _f^{ss^{\prime }}(\vec k,\tau
_1-\tau _2)\overline{\eta }_{s^{\prime }}(\vec k,\tau _2)]\}
\end{array}
\eqnum{142}
\end{equation}
where 
\begin{equation}
\Delta _f^{ss^{\prime }}(\vec k,\tau _1-\tau _2)=\delta ^{ss^{\prime
}}\Delta _f(\vec k,\tau _1-\tau _2)  \eqnum{143}
\end{equation}
with 
\begin{equation}
\Delta _f(\vec k,\tau _1-\tau _2)=\frac 12[\overline{n}_f(\vec k%
)e^{-\varepsilon (\vec k)\left| \tau _1-\tau _2\right| }-n_f(\vec k%
)e^{\varepsilon (\vec k)\left| \tau _1-\tau _2\right| }]  \eqnum{144}
\end{equation}
being the free fermion propagator in which 
\begin{equation}
\overline{n}_f(\vec k)=(1+e^{-\beta \varepsilon (\vec k)})^{-1},\text{ }n_f(%
\vec k)=(1+e^{\beta \varepsilon (\vec k)})^{-1}  \eqnum{145}
\end{equation}
are the familiar fermion distribution functions [4-6, 13].

Finally, let us calculate the generating functional $Z_g^0[\zeta ]$. From
the stationary condition $\delta I_g(c^{*},c,\overline{c}^{*},\overline{c}%
;\zeta ^{*},\zeta ,\overline{\zeta }^{*},\overline{\zeta })=0$ and the
boundary conditions: 
\begin{equation}
\overline{c}^{*}(\vec k,\beta )=\overline{c}^{*}(\vec k),\text{ }c^{*}(\vec k%
,\beta )=c^{*}(\vec k)\text{, }\overline{c}(\vec k,0)=\overline{c}(\vec k)%
\text{, }c(\vec k,0)=c(\vec k)  \eqnum{146}
\end{equation}
which are the same as those for scalar fields other than for fermion fields
[5], it is easy to derive from Eq. (110) or (112) the following equations of
motion 
\begin{equation}
\begin{array}{c}
\dot c(\vec k,\tau )+\omega (\vec k)c(\vec k,\tau )=-\zeta (\vec k,\tau ),
\\ 
\dot c^{*}(\vec k,\tau )-\omega (\vec k)c_a^{*}(\vec k,\tau )=\zeta _a^{*}(%
\vec k,\tau ), \\ 
\stackrel{.}{\overline{c}}(\vec k,\tau )+\omega (\vec k)\overline{c}_a(\vec k%
,\tau )=\overline{\zeta }_a(\vec k,\tau ), \\ 
\stackrel{\cdot }{\overline{c}^{*}}(\vec k,\tau )-\omega (\vec k)\overline{c}%
_a^{*}(\vec k,\tau )=-\overline{\zeta }_a^{*}(\vec k,\tau ).
\end{array}
\eqnum{147}
\end{equation}
By the same procedure as stated in Eqs. (115)-(120), the solutions to the
above equations can be found to be 
\begin{equation}
\begin{array}{c}
c(\vec k,\tau )=c(\vec k)e^{-\omega (\vec k)\tau }-\int_0^\tau d\tau
^{\prime }e^{-\omega (\vec k)(\tau -\tau ^{\prime })}\zeta (\vec k,\tau
^{\prime }), \\ 
c^{*}(\vec k,\tau )=c^{*}(\vec k)e^{\omega (\vec k)(\tau -\beta )}-\int_\tau
^\beta d\tau ^{\prime }e^{\omega (\vec k)(\tau -\tau ^{\prime })}\zeta ^{*}(%
\vec k,\tau ^{\prime }), \\ 
\overline{c}(\vec k,\tau )=\overline{c}(\vec k)e^{-\omega (\vec k)\tau
}+\int_0^\tau d\tau ^{\prime }e^{-\omega (\vec k)(\tau -\tau ^{\prime })}%
\overline{\zeta }(\vec k,\tau ^{\prime }), \\ 
\overline{c}^{*}(\vec k,\tau )=\overline{c}^{*}(\vec k)e^{\omega (\vec k%
)(\tau -\beta )}+\int_\tau ^\beta d\tau ^{\prime }e^{\omega (\vec k)(\tau
-\tau ^{\prime })}\overline{\zeta }^{*}(\vec k,\tau ^{\prime }).
\end{array}
\eqnum{148}
\end{equation}
Upon substituting the above solutions into Eq. (110) or Eq. (112), we find 
\begin{equation}
\begin{array}{c}
I_g^0(c^{*},c,\overline{c}^{*},\overline{c};\zeta ^{*},\zeta ,\overline{%
\zeta }^{*},\overline{\zeta })=\int d^3k\{e^{-\beta \omega (\vec k)}[%
\overline{c}^{*}(\vec k)\overline{c}(\vec k)-c^{*}(\vec k)c(\vec k)] \\ 
+e^{-\beta \omega (\vec k)}\int_0^\beta d\tau e^{\omega (\vec k)\tau }[c^{*}(%
\vec k)\zeta (\vec k,\tau )+\overline{c}^{*}(\vec k)\overline{\zeta }(\vec k%
,\tau )] \\ 
-\int_0^\beta d\tau e^{-\omega (\vec k)\tau }[\zeta ^{*}(\vec k,\tau )c(\vec 
k)+\overline{\zeta }^{*}(\vec k,\tau )\overline{c}(\vec k)]\} \\ 
-\int_0^\beta d\tau _1\int_0^\beta d\tau _2\int d^3k\theta (\tau _1-\tau
_2)e^{-\omega (\vec k)(\tau _1-\tau _2)}[\zeta ^{*}(\vec k,\tau _1)\zeta (%
\vec k,\tau _2) \\ 
-\overline{\zeta }^{*}(\vec k,\tau _1)\overline{\zeta }(\vec k,\tau _2)].
\end{array}
\eqnum{149}
\end{equation}
On inserting the above expression into the following integral given by the
stationary-phase method: 
\begin{equation}
Z_g^0[\zeta ]=\int D(\overline{c}^{*}\overline{c}cc^{*})\exp \{-\int d^3k[%
\overline{c}^{*}(\vec k)\overline{c}(\vec k)-c^{*}(\vec k)c(\vec k%
)]+I_g^0(c^{*},c,\overline{c}^{*},\overline{c};\zeta ^{*},\zeta ,\overline{%
\zeta }^{*},\overline{\zeta })\}  \eqnum{150}
\end{equation}
and applying the integration formulas in Eq. (39), one can get 
\begin{equation}
\begin{array}{c}
Z_g^0[\zeta ]=Z_g^0\exp \{\int_0^\beta d\tau _1\int_0^\beta d\tau _2\int
d^3k[\theta (\tau _1-\tau _2)-(1-e^{\beta \omega (\vec k)})^{-1}]e^{-\omega (%
\vec k)(\tau _1-\tau _2)} \\ 
\times [\overline{\zeta }^{*}(\vec k,\tau _1)\overline{\zeta }(\vec k,\tau
_2)-\zeta ^{*}(\vec k,\tau _1)\zeta (\vec k,\tau _2)]\}
\end{array}
\eqnum{151}
\end{equation}
where 
\begin{equation}
Z_g^0==\prod\limits_{\overrightarrow{k}}[1-e^{-\beta \omega (\vec k)}]^2 
\eqnum{152}
\end{equation}
is just the partition function arising from the free ghost particles which
plays the role of cancelling out the unphysical contribution contained in
Eq. (124). If we change the integration variables in Eq. (151) and
considering the relations 
\begin{equation}
\zeta ^{*}(\vec k,\tau )=-\overline{\zeta }(-\vec k,\tau )\text{, }\zeta (%
\vec k,\tau )=-\overline{\zeta }^{*}(-\vec k,\tau )  \eqnum{153}
\end{equation}
which will be interpreted in the next section, Eq. (151) may be recast in
the form 
\begin{equation}
\begin{array}{c}
Z_g^0[\zeta ]=Z_g^0\exp \{\int_0^\beta d\tau _1\int_0^\beta d\tau _2\int
d^3k[\overline{\zeta }^{*}(\vec k,\tau _1)\Delta _b(\vec k,\tau _1-\tau _2)%
\overline{\zeta }(\vec k,\tau _2) \\ 
-\zeta ^{*}(\vec k,\tau _1)\Delta _b(\vec k,\tau _1-\tau _2)\zeta (\vec k%
,\tau _2)]\}
\end{array}
\eqnum{154}
\end{equation}
where $\Delta _b(\vec k,\tau _1-\tau _2)$ was written in Eq. (128).

Up to the present, the perturbative expansion of the thermal QED generating
functional in the coherent-state representation has exactly been obtained by
the combination of Eqs. (99), (100), (123), (142) and (154). Especially, the
partition function for the free system has been given by the product of Eqs.
(124), (139) and (152). The partition function for the interacting system
can be calculated in the way as shown in Eq. (45). Here it should be noted
that the differential $\delta /\delta j(\tau )$ in Eq. (99) represents the
collection of the differentials $\delta /\delta \xi _\lambda ^{*}(\vec k%
,\tau )$, $\delta /\delta \xi _\lambda (\vec k,\tau )$, $-\delta /\delta
\eta _s(\vec k,\tau )$, $\delta /\delta \eta _s^{*}(\vec k,\tau )$, $-\delta
/\delta \overline{\eta }_s(\vec k,\tau )$, $\delta /\delta \overline{\eta }%
_s^{*}(\vec k,\tau )$, $-\delta /\delta \zeta (\vec k,\tau )$, $\delta
/\delta \zeta ^{*}(\vec k,\tau )$, $-\delta /\delta \overline{\zeta }(\vec k%
,\tau )$ and $\delta /\delta \overline{\zeta }^{*}(\vec k,\tau )$.

In the last part of this section, we briefly describe the generating
functional for the thermal $\varphi ^4$ theory. The generating functional
may be written out from Eq. (95) by letting $q=a.$ Its perturbative
expansion is still represented by Eq. (99) with noting that the $H_I(\frac 
\delta {\delta j(\tau )})$ is given by Eq. (94) with replacing $a_\alpha $
in Eq. (94) by $\delta /\delta j_\alpha ^{*}$ and 
\begin{equation}
\begin{array}{c}
Z_0[j]=\int D(a^{*}a)\exp \{-\int d^3ka^{*}(\vec k)a(\vec k)\}\int {\frak D}%
(a^{*}a)\exp \{I(a^{*},a;j^{*},j)]\} \\ 
=\int D(a^{*}a)\exp \{-\int d^3ka^{*}(\vec k)a(\vec k)+I_0(a^{*},a;j^{*},j)\}
\end{array}
\eqnum{155}
\end{equation}
where 
\begin{equation}
\begin{array}{c}
I(a^{*},a;j^{*},j)=\int d^3ka^{*}(\vec k,\beta )a(\vec k,\beta
)-\int_0^\beta d\tau \int d^3k\{a^{*}(\vec k,\tau )\dot a(\vec k,\tau ) \\ 
+\omega (\vec k)a^{*}(\vec k,\tau )a(\vec k,\tau )-j^{*}(\vec k,\tau )a(\vec 
k,\tau )-a^{*}(\vec k,\tau )j(\vec k,\tau )\}
\end{array}
\eqnum{156}
\end{equation}
and $I_0(a^{*},a;j^{*},j)$ is determined by the stationary condition $\delta
I(a^{*},a;j^{*},j)=0.$ Completely following the procedure described in Eqs.
(113)-(129) with noting the boundary condition 
\begin{equation}
a^{*}(\vec k,\beta )=a^{*}(\vec k)\text{, }a(\vec k,0)=a(\vec k)  \eqnum{157}
\end{equation}
and the relation 
\begin{equation}
j^{*}(\vec k,\tau )=j(-\vec k,\tau ),  \eqnum{158}
\end{equation}
which will be proved in the next section, it is easy to obtain 
\begin{equation}
Z_0[\xi ]=Z_b^0\exp \{\int_0^\beta d\tau _1\int_0^\beta d\tau _2\int
d^3kj^{*}(\vec k,\tau _1)\Delta _b(\vec k,\tau _1-\tau _2)j(\vec k,\tau _2)\}
\eqnum{159}
\end{equation}
where 
\begin{equation}
Z_b^0=\prod\limits_{\vec k}\frac 1{1-e^{-\beta \omega (\vec k)}}  \eqnum{160}
\end{equation}
and $\Delta _b(\vec k,\tau _1-\tau _2)$ was represented in Eq. (128).

\section{Derivation of the generating functional represented in the position
space}

In this section, we plan to show how the perturbative expansions of the
generating functionals represented in the position space are derived from
the corresponding ones given in the preceding section. For this purpose, we
need to derive the generating functional represented in position space for
free systems. For thermal QED, it can be written as 
\begin{equation}
Z^0[J]=Z_f^0[I,\overline{I}]Z_p^0[J_\mu ]Z_g^0[K,\overline{K}]  \eqnum{161}
\end{equation}
where $Z_f^0[I,\overline{I}]$, $Z_p^0[J_\mu ]$ and $Z_g^0[K,\overline{K}]$
are the position space generating functionals arising respectively from the
free fermions, photons and ghost particles and $I$, $\overline{I}$, $J_\mu $%
, $K$ and $\overline{K}$ are the sources coupled to the fermion, photon and
ghost particle fields, respectively. In order to write out the $Z_f^0[I,%
\overline{I}]$, $Z_p^0[J_\mu ]$ and $Z_g^0[K,\overline{K}]$ from the
generating functionals given in Eqs. (142), (123) and (154), it is necessary
to establish relations between the sources introduced in the position space
and the ones in the coherent-state representation. Let us separately discuss
the functionals $Z_f^0[I,\overline{I}]$, $Z_p^0[J_\mu ]$ and $Z_g^0[K,%
\overline{K}]$. First we focus our attention on the functioinal $Z_f^0[I,%
\overline{I}]$. Usually, the external source terms of fermions in the
generating functional given in the position space are of the form $%
\int_0^\beta d\tau \int d^3x[\overline{I}(\vec x,\tau )\psi (\vec x,\tau )+%
\overline{\psi }(\vec x,\tau )I(\vec x,\tau )]$ [5,6]. Substituting in this
form the Fourier expansions in Eq. (55) for the fermion field and in the
following for the sources 
\begin{equation}
\begin{array}{c}
I(\vec x,\tau )=\int \frac{d^3k}{(2\pi )^{3/2}}I(\vec k,\tau )e^{i\vec k%
\cdot \vec x} \\ 
\overline{I}(\vec x,\tau )=\int \frac{d^3k}{(2\pi )^{3/2}}\overline{I}(\vec k%
,\tau )e^{-i\vec k\cdot \vec x}
\end{array}
\eqnum{162}
\end{equation}
we have 
\begin{equation}
\begin{array}{c}
\int_0^\beta d\tau \int d^3x[\overline{I}(\vec x,\tau )\psi (\vec x,\tau )+%
\overline{\psi }(\vec x,\tau )I(\vec x,\tau )]=\int_0^\beta d\tau \int
d^3k[\eta _s^{*}(\vec k,\tau )b_s(\vec k,\tau ) \\ 
+b_s^{*}(\vec k,\tau )\eta _s(\vec k,\tau )+\overline{\eta }_s^{*}(\vec k%
,\tau )d_s(\vec k,\tau )+d_s^{*}(\vec k,\tau )\overline{\eta }_s(\vec k,\tau
)]
\end{array}
\eqnum{163}
\end{equation}
where 
\begin{equation}
\begin{array}{c}
\eta _s(,\tau )=\overline{u}_s(\vec k)I(\vec k,\tau )\text{, }\eta _s^{*}(%
\vec k,\tau )=\overline{I}(\vec k,\tau )u_s(\vec k), \\ 
\overline{\eta }_s(\vec k,\tau )=-\overline{I}(-\vec k,\tau )v_s(\vec k)%
\text{, }\overline{\eta }_s^{*}(\vec k,\tau )=-\overline{v}_s(\vec k)I(-\vec 
k,\tau ).
\end{array}
\eqnum{164}
\end{equation}
From these relations and the property of Dirac spinors, the relation in Eq.
(141) is easily obtained. On substituting Eq. (164) into Eq. (142), one can
get 
\begin{equation}
Z_f^0[I,\overline{I}]=Z_f^0\exp \{\int_0^\beta d\tau _1\int_0^\beta d\tau
_2\int d^3k\overline{I}(\vec k,\tau _1)S_F(\vec k,\tau _1-\tau _2)I(\vec k%
,\tau _2)\}  \eqnum{165}
\end{equation}
where 
\begin{equation}
S_F(\vec k,\tau _1-\tau _2)=[({\bf k}+M)/\varepsilon (\vec k)]\Delta _f(\vec 
k,\tau _1-\tau _2)  \eqnum{166}
\end{equation}
with ${\bf k=}\gamma _\mu k^\mu $ and $k^\mu =(\vec k,i\varepsilon _n)$. By
making use of the inverse transformation of Eq. (162), the generating
functional in Eq. (165) is finally represented as [5,6] 
\begin{equation}
Z_f^0[I,\overline{I}]=Z_f^0\exp \{\int_0^\beta d^4x_1\int_0^\beta d^4x_2%
\overline{I}(x_1)S_F(x_1-x_2)I(x_2)\}  \eqnum{167}
\end{equation}
where $x=(\vec x,\tau )$, $d^4x=d\tau d^3x$ and 
\begin{equation}
S_F(x_1-x_2)=\int \frac{d^3k}{(2\pi )^3}S_F(\vec k,\tau _1-\tau _2)e^{i\vec k%
\cdot \vec x}  \eqnum{168}
\end{equation}
It is well-known that the propagator $\Delta _f(\vec k,\tau _1-\tau _2)$ is
antiperiodic, 
\begin{equation}
\begin{array}{c}
\Delta _f(\vec k,\tau _1-\tau _2)=-\Delta _f(\vec k,\tau _1-\tau _2-\beta ),%
\text{ }if\text{ }\tau _1\succ \tau _2 \\ 
\Delta _f(\vec k,\tau _1-\tau _2)=-\Delta _f(\vec k,\tau _1-\tau _2+\beta ),%
\text{ }if\text{ }\tau _1\prec \tau _2
\end{array}
\eqnum{169}
\end{equation}
This can easily be proved from its representation in the operator formalism
shown in Eq. (46) with the help of the translation transformation $\widehat{b%
}_s(\tau )=e^{\beta \widehat{K}}$ $\widehat{b}_se^{-\beta \widehat{K}}$ and
the trace property $Tr(AB)=Tr(BA)$. According to the antiperiodic property
of the propagator, we have the following expansion 
\begin{equation}
\Delta _f(\vec k,\tau )=\frac 1\beta \sum_n\Delta _f(\vec k,\varepsilon
_n)e^{-i\varepsilon _n\tau }  \eqnum{170}
\end{equation}
where $\tau =\tau _1-\tau _2$, $\varepsilon _n=\frac \pi \beta (2n+1)$ with $%
n$ being the $%
\mathop{\rm integer}
$ and 
\begin{equation}
\Delta _f(\vec k,\varepsilon _n)=\int_0^\beta d\tau e^{i\varepsilon _n\tau
}\Delta _f(\vec k,\tau )=\frac{\varepsilon (\vec k)}{\varepsilon
_n^2+\varepsilon (\vec k)^2}  \eqnum{171}
\end{equation}
where the expression in Eq. (144) has been used. Substituting Eqs. (170)
into Eq. (166) and noticing the above expression, the propagator in Eq.
(168) can be written as [5,6] 
\begin{equation}
S_F(x_1-x_2)=\frac 1\beta \sum_n\int \frac{d^3k}{(2\pi )^3}\frac{e^{i%
\overrightarrow{k}\cdot (\vec x_1-\vec x_2)-i\varepsilon _n(\tau _1-\tau _2)}%
}{\vec \gamma \cdot \vec k+M-i\varepsilon _n\gamma ^0}  \eqnum{172}
\end{equation}

Next, we discuss the generating functional $Z_p^0[J_\mu ].$ The source term
of photons in the generating functional given in the position space is
commonly taken as $\int_0^\beta d\tau \int d^3xJ_\mu (\vec x,\tau )A^\mu (%
\vec x,\tau )$ [5,6]$.$ Employing the expansions in Eq. (56) and in the
following 
\begin{equation}
J_\mu (\vec x,\tau )=\int \frac{d^3k}{(2\pi )^{3/2}}J_\mu (\vec k,\tau )e^{i%
\vec k\cdot \vec x},  \eqnum{173}
\end{equation}
we can write 
\begin{equation}
\int_0^\beta d\tau \int d^3xJ_\mu (\vec x,\tau )A^\mu (\vec x,\tau
)=\int_0^\beta d\tau \int d^3k[\xi _\lambda ^{*}(\vec k,\tau )a_\lambda (%
\vec k,\tau )+a_\lambda ^{*}(\vec k,\tau )\xi _\lambda (\vec k,\tau )] 
\eqnum{174}
\end{equation}
where 
\begin{equation}
\xi _\lambda (\vec k,\tau )=(2\omega (\vec k))^{-1/2}\epsilon _\lambda ^\mu (%
\vec k)J_\mu (\vec k,\tau )=\xi _\lambda ^{*}(-\vec k,\tau )  \eqnum{175}
\end{equation}
in which the last equality follows from that the $J_\mu (\vec x,\tau )$ is a
real function. Inserting the relation in Eq. (175) and then the inverse
transformation of Eq. (173) into Eq. (123) and considering completeness of
the polarization vectors, one may find the generating functional $%
Z_p^0[J_\mu ]$ such that [5,6] 
\begin{equation}
Z_p^0[J_\mu ]=Z_p^0\exp \{-\frac 12\int_0^\beta d^4x_1\int_0^\beta
d^4x_2J^\mu (x_1)D_{\mu \nu }(x_1-x_2)J^\nu (x_2)\}  \eqnum{176}
\end{equation}
where 
\begin{equation}
D_{\mu \nu }(x_1-x_2)=g_{\mu \nu }\int \frac{d^3k}{(2\pi )^3}\frac 1{\omega (%
\vec k)}\Delta _b(\vec k,\tau _1-\tau _2)e^{i\vec k\cdot (\vec x_1-\vec x_2)}
\eqnum{177}
\end{equation}
By the same argument as mentioned for Eq. (169), it can be proved that the
photon propagator $\Delta _b(\vec k,\tau _1-\tau _2)$ is a periodic function 
\begin{equation}
\begin{array}{c}
\Delta _b(\vec k,\tau _1-\tau _2)=\Delta _b(\vec k,\tau _1-\tau _2-\beta ),%
\text{ }if\text{ }\tau _1\succ \tau _2; \\ 
\Delta _b(\vec k,\tau _1-\tau _2)=\Delta _b(\vec k,\tau _1-\tau _2+\beta ),%
\text{ }if\text{ }\tau _1\prec \tau _2.
\end{array}
\eqnum{178}
\end{equation}
Therefore, we have the expansion 
\begin{equation}
\Delta _b(\vec k,\tau )=\frac 1\beta \sum_n\Delta _b(\vec k,\omega
_n)e^{-i\omega _n\tau }  \eqnum{179}
\end{equation}
where $\omega _n=\frac{2\pi n}\beta $ and 
\begin{equation}
\Delta _b(\vec k,\omega _n)=\int_0^\beta d\tau e^{i\omega _n\tau }\Delta _b(%
\vec k,\tau )=\frac{\omega (\vec k)}{\varepsilon _n^2+\omega (\vec k)^2}. 
\eqnum{180}
\end{equation}
where the expression in Eq. (128) has been employed. Upon substituting Eqs.
(179) and (180) in Eq. (177), we arrive at [5,6] 
\begin{equation}
D_{\mu \nu }(x_1-x_2)=\frac 1\beta \sum_n\int \frac{d^3k}{(2\pi )^3}\frac{%
g_{\mu \nu }}{\varepsilon _n^2+\omega (\vec k)^2}e^{i\vec k\cdot (\vec x_1-%
\vec x_2)-i\omega _n(\tau _1-\tau _2)}  \eqnum{181}
\end{equation}
which just is the photon propagator given in the position space and in the
Feynman gauge.

Finally, we turn to the generating functional $Z_g^0[K,\overline{K}]$. In
accordance with the expansions in Eq. (57) for the ghost particle fields and
those for the external sources: 
\begin{equation}
\begin{array}{c}
K(\vec x,\tau )=\int \frac{d^3k}{(2\pi )^{3/2}}K(\vec k,\tau )e^{i\vec k%
\cdot \vec x} \\ 
\overline{K}(\vec x,\tau )=\int \frac{d^3k}{(2\pi )^{3/2}}\overline{K}(\vec k%
,\tau )e^{-i\vec k\cdot \vec x}
\end{array}
\eqnum{182}
\end{equation}
the relation between the sources in the position space and the ones in the
coherent-state representation can be found to be 
\begin{equation}
\begin{array}{c}
\int_0^\beta d\tau \int d^3x[\overline{K}(\vec x,\tau )C(\vec x,\tau )+%
\overline{C}(\vec x,\tau )K(\vec x,\tau )]=\int_0^\beta d\tau \int
d^3k[\zeta ^{*}(\vec k,\tau )c(\vec k,\tau ) \\ 
+c^{*}(\vec k,\tau )\zeta (\vec k,\tau )+\overline{\zeta }^{*}(\vec k,\tau )%
\overline{c}(\vec k,\tau )+\overline{c}^{*}(\vec k,\tau )\overline{\zeta }(%
\vec k,\tau )]
\end{array}
\eqnum{183}
\end{equation}
where 
\begin{equation}
\begin{array}{c}
\zeta (\vec k,\tau )=(2\omega (\vec k))^{-1/2}K(\vec k,\tau ),\text{ }\zeta
^{*}(\vec k,\tau )=(2\omega (\vec k))^{-1/2}\overline{K}(\vec k,\tau ), \\ 
\overline{\zeta }(\vec k,\tau )=-(2\omega (\vec k))^{-1/2}\overline{K}(-\vec 
k,\tau ),\text{ }\overline{\zeta }^{*}(\vec k,\tau )=-(2\omega (\vec k%
))^{-1/2}K(-\vec k,\tau ).
\end{array}
\eqnum{184}
\end{equation}
from which the relations denoted in Eq. (153) can directly be deduced. When
the above relations and the inverse transformations of Eq. (182) are
inserted into Eq. (154), one can get 
\begin{equation}
Z_g^0[K,\overline{K}]=Z_g^0\exp \{-\int_0^\beta d^4x_1\int_0^\beta d^4x_2%
\overline{K}(x_1)\Delta _g(x_1-x_2)K(x_2)\}  \eqnum{185}
\end{equation}
where 
\begin{equation}
\begin{array}{c}
\Delta _g(x_1-x_2)=\int \frac{d^3k}{(2\pi )^3}\frac 1{\omega (%
\overrightarrow{k})}\Delta _b(\vec k,\tau _1-\tau _2)e^{i\vec k\cdot (\vec x%
_1-\vec x_2)} \\ 
=\frac 1\beta \sum\limits_n\int \frac{d^3k}{(2\pi )^3}\frac 1{\varepsilon
_n^2+\omega (\overrightarrow{k})^2}e^{i\vec k\cdot (\vec x_1-\vec x%
_2)-i\omega _n(\tau _1-\tau _2)}
\end{array}
\eqnum{186}
\end{equation}
which just is the free ghost particle propagator given in the position space
[5,6]. In the derivation of the last equality of Eq. (186), the expansion
given in Eqs. (179) and (180) have been used.

With the generating functionals given in Eqs. (167), (176) and (185), the
zeroth-order generating functional in Eq. (161) is explicitly represented in
terms of the propagators and external sources. Clearly, the exact generating
functional can immediately be written out from Eq. (99) as shown in the
following 
\begin{equation}
Z[J]=\exp \{-\int_0^\beta d^4x{\cal H}_I(\frac \delta {\delta J(x)})\}Z^0[J]
\eqnum{187}
\end{equation}
where $J$ stands for $I$, $\overline{I}$, $J_\mu $ $K$ and $\overline{K}$, $%
\frac \delta {\delta J(x)}$ represents the differentials $\frac \delta {%
\delta \overline{I}(x)}$, $-\frac \delta {\delta I(x)}$, $\frac \delta {%
\delta J_\mu (x)}$, $\frac \delta {\delta \overline{K}(x)}$ and $-\frac 
\delta {\delta K(x)}$ and ${\cal H}_I(\frac \delta {\delta J(x)})$ can be
written out from the last term in Eq. (54) when the field functions in the
term are replaced by the differentials with respect to the corresponding
sources.

At last, we sketch the derivation of the generating functional given in the
position space for the thermal $\varphi ^4$ theory. By using the expansions
in Eq. (85) for the scalar field and in the following for the source, 
\begin{equation}
J(\vec x,\tau )=\int \frac{d^3k}{(2\pi )^{3/2}}J(\vec k,\tau )e^{i\vec k%
\cdot \vec x},  \eqnum{188}
\end{equation}
it is easy to find the relation between the both sources given in the
position space and the coherent-state representation 
\begin{equation}
\int_0^\beta d\tau \int d^3xJ(\vec x,\tau )\varphi (\vec x,\tau
)=\int_0^\beta d\tau \int d^3k[j^{*}(\vec k,\tau )a(\vec k,\tau )+a^{*}(\vec 
k,\tau )j(\vec k,\tau )]  \eqnum{189}
\end{equation}
where 
\begin{equation}
j(\vec k,\tau )=(2\omega (\vec k))^{-1/2}J(\vec k,\tau )=(2\omega (\vec k%
))^{-1/2}J^{*}(-\vec k,\tau )=j^{*}(-\vec k,\tau )  \eqnum{190}
\end{equation}
The second equality in the above follows from the $%
\mathop{\rm real}
$ character of the source $J(\vec x,\tau )$. When the above relation and the
inverse transformation of Eq. (188) are inserted into Eq. (159), the
generating functional $Z_g^0[J]$ is found to be [5,6] 
\begin{equation}
Z_g^0[J]=Z_g^0\exp \{\frac 12\int_0^\beta d^4x_1\int_0^\beta
d^4x_2J(x_1)\Delta (x_1-x_2)J(x_2)\}  \eqnum{191}
\end{equation}
in which 
\begin{equation}
\begin{array}{c}
\Delta (x_1-x_2)=\int \frac{d^3k}{(2\pi )^3}\frac 1{\omega (\vec k)}\Delta
_b(\vec k,\tau _1-\tau _2)e^{i\vec k\cdot (\vec x_1-\vec x_2)} \\ 
=\frac 1\beta \sum_n\int \frac{d^3k}{(2\pi )^3}\frac 1{\varepsilon
_n^2+\omega (\vec k)^2}e^{i\vec k\cdot (\vec x_1-\vec x_2)-i\omega _n(\tau
_1-\tau _2)}
\end{array}
\eqnum{192}
\end{equation}
where the expressions in Eqs. (179) and (180) have been considered. The
exact generating functional is still represented by Eq. (187), but the $%
{\cal H}_I(\frac \delta {\delta J(x)})$ in Eq. (187) is now given by the
last term in Eq. (84) with replacing $\varphi (x)$ by $\frac \delta {\delta
J(x)}$.

\section{Concluding remarks}

In this paper, the path-integral formalism of the thermal QED and $\varphi
^4 $ theory has been correctly established in the coherent-state
representation. In contrast to the ordinary path-integral formalism set up
in the position space, the expressions of the generating functionals
presented in this paper not only gives an alternative quantization for these
theories, but also provides an alternative and general method of calculating
the partition functions, the thermal Green's functions and thereby other
statistical quantities in the coherent-state representation. In particular,
the generating functional enables us to carry out analytical calculations
without concerning its original discretized form. The discretized form must
be used for the generating functionals given in the previous literature
because the previous generating functionals given in the coherent-state
representation is incorrect and actually useless. As one has seen from Sect.
4, the analytical calculations of the zero-order generating functional is
more simple and direct than the previous calculations performed in the
discretized form either in the coherent-state representation or in the
position space [4-6]. The coherent-state path-integral formalism established
in the coherent-state representation corresponds to the operator formalism
formulated in terms of creation and annihilation operators. In comparison
with the operator formalism which was widely applied in the many-body theory
[4,13], the coherent-state path-integral formalism has a prominent advantage
that in the calculations within this formalism, use of the operator
commutators and the Wick theorem is completely avoided. Therefore, it is
more convenient for practical applications. It should be mentioned that
although the QED generating functional is derived in the Feynman gauge, the
result is exact. This is because QED is a gauge-independent theory. As shown
in Sect. 4, in the partition function shown in Eq. (124) which is derived in
the Feynman gauge, the contribution given by the unphysical degrees of
freedom, i.e., the time and longitudinal polarizations of photons is
completely cancelled out by the partition function shown in Eq. (152) which
is contributed from the unphysical ghost particles. This indicates the
necessity of retaining the ghost term in the effective Lagrangian.
Certainly, the generating functional formulated in the coherent-state
representation can be established in arbitrary gauges. But, in this case,
the photon propagator would have a rather complicated form due to that the
longitudinal part of the propagator will involve the photon polarization
vector. Another point we would like to mention is that to formulate the
quantization of the thermal QED and $\varphi ^4$ theory in the
coherent-state representation, we limit ourself to work in the
imaginary-time formalism. It is no doubt that the theory can equally be
described in the real-time formalism. Needless to say, the coherent-state
formalism described in this paper can readily be extended to other fields
such as the quantum chromodynamics (QCD) and the quantum hadrondynamics
(QHD). Discussions on these subjects and practical applications of the
coherent-state formalism described in this paper will be anticipated in the
future.

\section{Acknowledgment}

This work was supported by National Natural Science Foundation of China.

\section{\bf Reference}

[1] A. Casher, D. Lurie and M. Revzen, J. Math. Phys. {\bf 9}, 1312 (1968).

[2] J. S. Langer, Phys. Rev. {\bf 167}, 183 (1968).

[3] L. S. Schulman, Techniques and Applications of Path Integration, A
wiley-Interscience

Publication, New York (1981).

[4] J. W. Negele and H. Orland, Quantum Many-Particle Systems, Perseus Books
Publishing, L.L.C, 1998.

[5] M. Le Bellac, Thermal Field Theory, Cambridge University Press, 1996.

[6] J. I. Kapusta, Finite-Temperature Field Theory, Cambridge University
Press, 1989

[7] J. C. Su, Phys. Lett. A {\bf 268}, 279 (2000).

[8] S. S. Schweber, J. Math, Phys. {\bf 3}, 831 (1962).

[9] L. D. Faddeev and A. A. Slavnov, Gauge Fields: Introduction to Quantum
Theory,

The Benjamin/Cummings Publishing Company, Inc. (1980).

[10] C. Itzykson and J-B. Zuber, Quantum Field Theory, McGraw-Hill Inc. New
York (1980).

[11] J. R. Klauder, Phys. Rev. D {\bf 19}, 2349 (1979).

.

[12] R. J. Glauber, Phys. Rev. {\bf 131}, 2766 (1963).

[13] A. L. Fetter and J. D. Walecka, Quantu Quantum Theory of Many-Particle
System, McGraw-Hill

(1971).

[14] D. Lurie, Particles and Fields, Interscience Publishers, a divison of
John Viley \& Sons, New York, 1968.

\end{document}